\documentclass[prd,nopacs,preprintnumbers,twocolumn,amsmath,nofootinbib,amssymb]{revtex4}
\usepackage{graphicx,color,dcolumn,booktabs,bm}
\usepackage{longtable,lscape}
\usepackage{txfonts}
\usepackage{overpic}
\usepackage{epstopdf}
\usepackage{indentfirst}
\usepackage{feynmf}   
\usepackage{slashed}  
\usepackage{cases}
\usepackage{color}
\usepackage{float}
\usepackage{makecell}
\usepackage{multirow}
\usepackage{graphicx,color,dcolumn,booktabs,bm}
\usepackage{epsfig,dsfont,amssymb,amsmath,amsfonts,amsbsy,mathrsfs}

\graphicspath{{Figures/}} %

\usepackage{hyperref}
\hypersetup{colorlinks,citecolor=blue,anchorcolor=red,menucolor=red, linkcolor=red,filecolor=red,runcolor=red,urlcolor=blue,frenchlinks=red}

\makeatletter
\@addtoreset{equation}{section}
\makeatother

\allowdisplaybreaks

\newcolumntype{I}{!{\vrule width 0.9pt}}

\begin{document}

\title{Restudy of the color-allowed two-body nonleptonic decays of bottom baryons ${\Xi_b}$ and ${\Omega_b}$ supported by hadron spectroscopy}
\author{Yu-Shuai Li$^{1,2}$}\email{liysh20@lzu.edu.cn}
\author{Xiang Liu$^{1,2,3}$\footnote{Corresponding author}}\email{xiangliu@lzu.edu.cn}
\affiliation{$^1$School of Physical Science and Technology, Lanzhou University, Lanzhou 730000, China\\
$^2$Research Center for Hadron and CSR Physics, Lanzhou University and Institute of Modern Physics of CAS, Lanzhou 730000, China\\
$^3$Lanzhou Center for Theoretical Physics, Key Laboratory of Theoretical Physics of Gansu Province, and Frontier Science Center for Rare Isotopes, Lanzhou University, Lanzhou 730000, China}

\begin{abstract}
In this work, we calculate the branching ratios of the color-allowed two-body nonleptonic decays of the bottom baryons, which include the $\Xi_b\to \Xi_c^{(*)}$ and $\Omega_b\to \Omega_c^{(*)}$ weak transitions by emitting a pseudoscalar meson ($\pi^{-}$, $K^{-}$, $D^{-}$, and $D_s^{-}$) or a vector meson ($\rho^{-}$, $K^{*-}$, $D^{*-}$, and $D_s^{*-}$). For achieving this aim, we adopt the three-body light-front quark model with the support of hadron spectroscopy, where the spatial wave functions of these heavy baryons involved in these weak decays are obtained by a semirelativistic potential model associated with the Gaussian expansion method. Our results show that these decays with the $\pi^-$, $\rho^-$, and $D_s^{(*)-}$-emitted mode have considerable widths, which could be accessible at the ongoing LHCb and Belle II experiments.
\end{abstract}

\maketitle

\section{introduction}
\label{sec1}

The investigation of bottom baryon weak decay has aroused the attentions from both theorist and experimentalist. It is not only an important approach to deepen our understanding to the dynamics of the weak transition, but also is the crucial step of searching for new physics beyond the Standard Model (SM).

Taking this opportunity, we want to introduce several recent progresses. As we know, the lepton flavor universality (LFU) violation has been examined in various $b\to c$ weak transitions \cite{BaBar:2012obs,BaBar:2013mob,Belle:2015qfa,LHCb:2015gmp,Belle:2016dyj,Belle:2019rba,FermilabLattice:2021cdg} in the past decade. The measurement of the ratio $R_{D^{(*)}}=\mathcal{B}(B\to D^{(*)}\tau\nu_\tau)/\mathcal{B}(B\to D^{(*)}e(\mu)\nu_{e(\mu)})$ \cite{BaBar:2012obs,BaBar:2013mob,Belle:2015qfa,LHCb:2015gmp,Belle:2016dyj,Belle:2019rba,FermilabLattice:2021cdg} shows the discrepancy with the prediction of the SM \cite{HFLAV:2019otj}, which indicates the possible evidence of new physics. Inspired by the anomalies of $R_{D^{(*)}}$ existing in the $b\to c$ weak transitions, it is interesting to study the corresponding ratios for the bottom baryon weak decays like $\Xi_b\to\Xi_c\ell^{-}\nu_{\ell}$ and $\Omega_b\to\Omega_c\ell^{-}\nu_{\ell}$, where the key point is to calculate the form factors involved in the corresponding  weak transition of the bottom baryon into the charmed baryon.
For the nonleptonic decays of the bottom baryon, a series of intriguing measurements were performed,
which include the observation of charmful and charmless modes \cite{CDF:2008llm,LHCb:2014yin,LHCb:2016rja,ParticleDataGroup:2020ssz}, the discovery of the hidden-charm pentaquark states $P_{c}(4312)$, $P_{c}(4380)$, $P_{c}(4440)$, and $P_{c}(4457)$ in the $\Lambda_b\to J/\psi pK$ process \cite{LHCb:2015yax,LHCb:2019kea}, and $P_{cs}(4459)$ in the $\Xi_b\to J/\psi\Lambda K$ process \cite{LHCb:2020jpq}. 
These efforts make us gain a deeper understanding of the dynamics involved in the heavy-flavor baryon weak decays.

Although great progress had been made, continuing to explore
new allowed decay modes of the bottom baryons is a research issue full of opportunity [see the Particle Data Group (PDG) \cite{ParticleDataGroup:2020ssz} for learning the present experimental status]. 
With the accumulation of experimental data, the LHCb experiment shows its potential to explore the allowed decays of the bottom baryons like the $\Xi_b$ and $\Omega_b$ states,
which is still missing in the PDG.
Besides, with the KEKB upgrading to the SuperKEKB, the center-of-mass energy of the $e^+e^-$ collision may reach up to 11.24 GeV. The ongoing Belle II \cite{Belle-II:2018jsg} should be a potential experiment to perform the study on the bottom-flavor physics. Facing this exciting status, we have reason to believe that it is suitable time to investigate the two-body nonleptonic decays of the $\Xi_b$ and $\Omega_b$ baryons, which is the main task of this work.

The bottom baryon weak decays have been widely studied by various approaches including the quark models \cite{Cheng:1996cs,Ivanov:1997hi,Ivanov:1997ra,Albertus:2004wj,Ebert:2006rp,Gutsche:2018utw,Faustov:2018ahb,Geng:2020ofy}, the flavor symmetry method \cite{Zhao:2018zcb}, the light-front approach \cite{Chua:2018lfa,Chua:2019yqh,Ke:2019smy,Zhu:2018jet,Zhao:2018zcb,Li:2021qod}, and the quantum chromodynamics (QCD) sum rules \cite{Wang:2008sm,Khodjamirian:2011jp,Wang:2015ndk,Zhao:2020mod}. For these theoretical studies, how to estimate the form factors of the weak transition is the key issue. Additionally, for the bottom baryon weak decays, how to optimize the three-body problem is also a challenge. Usually, the quark-diquark scheme as an approximate treatment was widely used in previous theoretical works \cite{Guo:2005qa,Zhu:2018jet,Zhao:2018zcb,Chua:2018lfa,Chua:2019yqh,Ke:2019smy}.
And the spatial wave functions of these hadrons involved in the bottom baryon weak decays
are approximately taken as a simple harmonic oscillator wave function, which makes the results dependent on the parameter of the harmonic oscillator wave function.
For avoiding the uncertainty from these approximate treatments mentioned above,
in this work we calculate the weak transition form factors of the $\Xi_b\to\Xi_c^{(*)}$ and $\Omega_b\to\Omega_c^{(*)}$ transitions with emitting a pseudoscalar meson ($\pi^{-}$, $K^{-}$, $D^{-}$ and $D_s^{-}$) or a vector meson ($\rho^{-}$, $K^{*-}$, $D^{*-}$, and $D_s^{*-}$) in the three-body light-front quark model. Here, $\Xi_c^{(*)}$ denotes the ground state $\Xi_c$ or its first radial excited state $\Xi_c(2970)$, while $\Omega_c^{(*)}$ represents the ground state $\Omega_c$ or its first radial excited state $\Omega_c(2S)$. In the realistic calculation, we take the numerical spatial wave functions of these involved bottom and charmed baryons as input, 
where the semirelativistic potential model \cite{Capstick:1985xss,Li:2021qod} associated with the Gaussian expansion method (GEM) \cite{Hiyama:2003cu,Yoshida:2015tia,Hiyama:2018ivm,Yang:2019lsg} is adopted. 
By fitting the mass spectrum of these observed bottom and charmed baryons, the parameters of the adopted semirelativistic potential model can be fixed. 
Comparing with former approximation of taking a simple harmonic oscillator wave function \cite{Guo:2005qa,Zhu:2018jet,Zhao:2018zcb,Chua:2018lfa,Chua:2019yqh,Ke:2019smy}, the treatment given in this work can avoid the uncertainties resulting from the selection of the spatial wave function of the heavy baryon. 
Thus, the color-allowed two-body nonleptonic decays of bottom baryons ${\Xi_b}$ and ${\Omega_b}$ with the support of hadron spectroscopy as a development. In the following sections, more details will be illustrated.

This paper is organized as follows. After the introduction, the formula of the form factors of the weak transitions $\Xi_b\to\Xi_c^{(*)}$ and $\Omega_b\to\Omega_c^{(*)}$ is given in Sec. \ref{sec2}. For getting the numerical spatial wave functions of these involved heavy baryons, we introduce the adopted semirelativistic potential model and GEM. With these results as input, the calculated concerned form factors are displayed. In Sec. \ref{sec3}, we study the color-allowed two-body nonleptonic decays with emitting a pseudoscalar meson ($\pi^{-}$, $K^{-}$, $D^{-}$, and $D_s^{-}$) or vector meson ($\rho^{-}$, $K^{*-}$, $D^{*-}$, and $D_s^{*-}$) in the na\"ive factorization assumption. Finally, the paper ends with a short summary.

\section{The transition form factors of the bottom baryon to the charmed baryon}
\label{sec2}

In this section, we briefly introduce how to calculate the form factors discussed in this work. Given that the quarks are confined in hadron, the weak transition matrix element cannot be calculated in the framework of perturbative QCD. Usually, the weak transition matrix element can be parametrized in terms of a series of dimensionless form factors \cite{Li:2021qod,Ke:2019smy}
\begin{equation}
\begin{split}
\langle&\mathcal{B}_{c}(1/2^{+})(P^{\prime},J^{\prime}_z)|\bar{c}\gamma^{\mu}(1-\gamma_5)b|\mathcal{B}_b(1/2^+)(P,J_z)\rangle\\
&=\bar{u}(P',J'_z)\left[f^V_1(q^2)\gamma^{\mu}+i\frac{f^V_2(q^2)}{M}\sigma^{\mu\nu}q_{\nu}+\frac{f^V_3(q^2)}{M}q^{\mu}\right.\\
&\quad\left.-\left(g^A_1(q^2)\gamma^{\mu}+i\frac{g^A_2(q^2)}{M}\sigma^{\mu\nu}q_{\nu}+\frac{g^A_3(q^2)}{M}q^{\mu}\right)\gamma_5\right]u(P,J_z)
\label{eq:formfactors}
\end{split}
\end{equation}
for the transitions of the bottom baryon to the charmed baryon. Here, $M(P)$ and $M'(P')$ are the mass(four-momentum) for the initial and final baryons, respectively, $\sigma^{\mu\nu}=i[\gamma^{\mu},\gamma^{\nu}]/2$, and $q=P-P'$ denotes the transferred momentum between the initial and final baryons.

The vertex function of a single heavy-flavor baryon $\mathcal{B}_{Q}$ ($Q=b,c$) with the spin $J=1/2$ and the  momentum $P$ is
\begin{equation}
\begin{split}
|\mathcal{B}_{Q}&(P,J,J_z)\rangle=\int\frac{d^3\tilde{p}_{1}}{2(2\pi)^3}\frac{d^3\tilde{p}_{2}}{2(2\pi)^3}\frac{d^3\tilde{p}_{3}}{2(2\pi)^3}2(2\pi)^3\\
&\times\sum_{\lambda_1,\lambda_2,\lambda_3}\Psi^{J,J_z}(\tilde{p}_i,\lambda_i)C^{\alpha\beta\gamma}
\delta^{3}(\tilde{\bar{P}}-\tilde{p}_1-\tilde{p}_2-\tilde{p}_3)\\
&\times~F_{nnQ}~|n_{\alpha}(\tilde{p}_1,\lambda_1)\rangle~|n_{\beta}(\tilde{p}_2,\lambda_2)\rangle~|Q_{\gamma}(\tilde{p}_3,\lambda_3)\rangle.
\end{split}
\end{equation}
Here, $n=u,d,s$ is the light-flavor quark, $C^{\alpha\beta\gamma}$ and $F_{nnQ}$ represent the color and flavor factors, and $\lambda_i$ and $p_i$ ($i$=1,2,3) are the helicities and light-front momenta of the on-mass-shell quarks, respectively, defined as
\begin{equation}
\tilde{p}_i=(p_i^+, p_{i\bot}), \quad p_i^+=p_i^0+p_i^3, \quad p_{i\bot}=(p_i^1, p_i^2).
\end{equation}

As suggested in Ref. \cite{Tawfiq:1998nk}, the spin and spatial wave functions for $\mathcal{B}_{Q}(\bar{3}_f)$ and $\mathcal{B}_{Q}(6_f)$ with the spin-parity quantum number  $J^{P}=1/2^{+}$ are written as
\begin{equation}
\begin{split}
\Psi^{J,J_z}(\tilde{p}_i,\lambda_i)=&A_0\bar{u}(p_1,\lambda_1)[(\slashed{\bar{P}}+M_0)\gamma_5]v(p_2,\lambda_2)\\
&\times\bar{u}_{Q}(p_3,\lambda_3)u(\bar{P},J,J_z)\phi(x_i,k_{i\bot}),\\
\Psi^{J,J_z}(\tilde{p}_i,\lambda_i)=&A_1\bar{u}(p_1,\lambda_1)[(\slashed{\bar{P}}+M_0)\gamma_{\bot\alpha}]v(p_2,\lambda_2)\\
&\times\bar{u}_{Q}(p_3,\lambda_3)\gamma_{\bot}^{\alpha}\gamma_5u(\bar{P},J,J_z)\phi(x_i,k_{i\bot})
\end{split}
\end{equation}
with $$A_0=\sqrt{3}A_1=\frac{1}{\sqrt{16\bar{P}^+M_0^3(e_1+m_1)(e_2+m_2)(e_3+m_3)}}$$  representing the normalization factor \cite{Ke:2019smy}.

In the framework of the three-body light-front quark model, the general expressions are written as \cite{Li:2021qod,Ke:2019smy}
\begin{widetext}
\begin{equation}
\begin{split}
\langle\mathcal{B}&_{c}^{(*)}\left(\bar{3}_f,1/2^+\right)(\bar{P}',J'_z)|\bar{c}\gamma^{\mu}(1-\gamma_5)b|\mathcal{B}_b\left(\bar{3}_f,1/2^+\right)(\bar{P},J_z)\rangle\\
=&\int\left(\frac{dx_1 d^{2}\vec{k}_{1\bot}}{2(2\pi)^3}\right)\left(\frac{dx_2 d^{2}\vec{k}_{2\bot}}{{2(2\pi)^3}}\right)
\frac{\phi(x_i,\vec{k}_{i\bot})\phi^{*}(x_i',\vec{k}_{i\bot}')}{16\sqrt{x_3 x_3' {M_0}^3 {M_0'}^3}}
\frac{\text{Tr}[(\slashed{\bar{P}}'-M_0')\gamma_5(\slashed{p}_1+m_1)(\slashed{\bar{P}}+M_0)\gamma_5(\slashed{p}_2-m_2)]}{\sqrt{(e_1+m_1) (e_2+m_2)(e_3+m_3)(e_1'+m_1')(e_2'+m_2')(e_3'+m_3')}}\\
&\times\bar{u}(\bar{P}',J_z')(\slashed{p}_3'+m_3')\gamma^{\mu}(1-\gamma_5)(\slashed{p}_3+m_3)u(\bar{P},J_z),
\label{eq:LFformfactor01}
\end{split}
\end{equation}
\begin{equation}
\begin{split}
\langle\mathcal{B}&_{c}^{(*)}\left(6_f,1/2^+\right)(\bar{P}',J'_z)|\bar{c}\gamma^{\mu}(1-\gamma_5)b|\mathcal{B}_b\left(6_f,1/2^+\right)(\bar{P},J_z)\rangle=\\
\int&\left(\frac{dx_1 d^{2}\vec{k}_{1\bot}}{2(2\pi)^3}\right)\left(\frac{dx_2 d^{2}\vec{k}_{2\bot}}{{2(2\pi)^3}}\right)
\frac{\phi(x_i,\vec{k}_{i\bot})\phi^{*}(x_i',\vec{k}_{i\bot}')}{48\sqrt{x_3 x_3' {M_0}^3 {M_0'}^3}}
\frac{\text{Tr}[\gamma_{\bot}^{\alpha}(\slashed{\bar{P}}'+M_0')(\slashed{p}_1+m_1)(\slashed{\bar{P}}+M_0)\gamma_{\bot}^{\beta}(\slashed{p}_2-m_2)]}{\sqrt{(e_1+m_1) (e_2+m_2)(e_3+m_3)(e_1'+m_1')(e_2'+m_2')(e_3'+m_3')}}\\
&\times\bar{u}(\bar{P}',J_z')\gamma_{\bot\alpha}\gamma_{5}(\slashed{p}_3'+m_3')\gamma^{\mu}(1-\gamma_5)(\slashed{p}_3+m_3)\gamma_{\bot\beta}\gamma_{5}u(\bar{P},J_z),
\label{eq:LFformfactor02}
\end{split}
\end{equation}
for the $\mathcal{B}_b(\bar{3}_f)\to\mathcal{B}_c(\bar{3}_f)$ and $\mathcal{B}_b(6_f)\to\mathcal{B}_c(6_f)$ transitions, respectively. Here, $\bar{P}=p_1+p_2+p_3$ and $\bar{P}'=p_1+p_2+p_3'$ are the light-front momenta for initial and final baryons, respectively, considering $p_1=p_1'$ and $p_2=p_2'$ in the spectator scheme, while $\phi$ and $\phi^{*}$ represent the spatial wave functions for the initial bottom baryon and the final charmed baryon, respectively. In the previous references \cite{Guo:2005qa,Zhu:2018jet,Zhao:2018zcb,Chua:2018lfa,Chua:2019yqh,Ke:2019smy}, the wave functions for baryon are usually treated as a simple harmonic oscillator forms with the oscillator parameter $\beta$, which results in
the $\beta$ dependence of the result. For avoiding this uncertainty, in this work, we adopt the numerical spatial wave functions for these involved baryons calculated by solving the three-body Schr\"{o}dinger equation with the semirelativistic quark model.

To calculate the form factors defined in Eq. (\ref{eq:formfactors}) from Eqs. (\ref{eq:LFformfactor01})-(\ref{eq:LFformfactor02}),  $V^{+}$, $A^{+}$, $\vec{q}_{\bot}\cdot\vec{V}$, $\vec{q}_{\bot}\cdot\vec{A}$, $\vec{n}_{\bot}\cdot\vec{V}$, and $\vec{n}_{\bot}\cdot\vec{A}$ are applied within a special gauge $q^{+}=0$. The details can be found in Ref. \cite{Chua:2019yqh}. Finally, the form factors are expressed as \cite{Li:2021qod}
\begin{equation}
\begin{split}
f^V_1(q^2)&=\int\mathcal{DS}_{0}\frac{1}{8\bar{P}^+\bar{P}'^+}
\text{Tr}[(\slashed{\bar{P}}+M_0)\gamma^+(\slashed{\bar{P}}'+M_0')(\slashed{p}_3'+m_3')\gamma^+(\slashed{p}_3+m_3)],\\
f^V_2(q^2)&=\int\mathcal{DS}_{0}\frac{iM}{8\bar{P}^+\bar{P}'^+q^2}
\text{Tr}[(\slashed{\bar{P}}+M_0)\sigma^{+\mu}q_{\mu}(\slashed{\bar{P}}'+M_0')(\slashed{p}_3'+m_3')\gamma^+(\slashed{p}_3+m_3)],\\
f^V_3(q^2)&=\frac{M}{M+M'}\left(f_1^V(q^2)(1-\frac{2\bar{P}\cdot q}{q^2})
+\int\frac{\mathcal{DS}_{0}}{4\sqrt{\bar{P}^+\bar{P}'^+}q^2}
\text{Tr}[(\slashed{\bar{P}}+M_0)\gamma^+(\slashed{\bar{P}}'+M_0')(\slashed{p}_3'+m_3')\slashed{q}\slashed{p}_3+m_3)]\right),\\
g^A_1(q^2)&=\int\mathcal{DS}_{0}\frac{1}{8\bar{P}^+\bar{P}'^+}
\text{Tr}[(\slashed{\bar{P}}+M_0)\gamma^+\gamma_5(\slashed{\bar{P}}'+M_0')(\slashed{p}_3'+m_3')\gamma^+\gamma_5(\slashed{p}_3+m_3)],\\
g^A_2(q^2)&=\int\mathcal{DS}_{0}\frac{-iM}{8\bar{P}^+\bar{P}'^+q^2}
\text{Tr}[(\slashed{\bar{P}}+M_0)\sigma^{+\mu}q_{\mu}\gamma_5(\slashed{\bar{P}}'+M_0')(\slashed{p}_3'+m_3')\gamma^+\gamma_5(\slashed{p}_3+m_3)],\\
g^A_3(q^2)&=\frac{M}{M-M'}\left(g^A_1(q^2)(-1+\frac{2\bar{P}\cdot q}{q^2})
+\int\frac{-\mathcal{DS}_{0}}{4\sqrt{\bar{P}^+\bar{P}'^+}q^2}
\text{Tr}[(\slashed{\bar{P}}+M_0)\gamma^+\gamma_5(\slashed{\bar{P}}'+M_0')(\slashed{p}_3'+m_3')~\slashed{q}~\gamma_5~(\slashed{p}_3+m_3)]\right),\\
\mathcal{DS}_{0}&=\frac{dx_1d^2\vec{k}_{1\bot}dx_2d^2\vec{k}_{2\bot}}{2(2\pi)^32(2\pi)^3}
\frac{\phi^*(x_i',\vec{k}_{i\bot}')\phi(x_i,\vec{k}_{i\bot})}{16\sqrt{x_3x_3'M_0^3M_0'^3}}
\frac{\text{Tr}[(\slashed{\bar{P}}'-M_0')\gamma_5(\slashed{p}_1+m_1)(\slashed{\bar{P}}+M_0)\gamma_5(\slashed{p}_2-m_2)]}{\sqrt{(e_1+m_1)(e_2+m_2)(e_3+m_3)(e_1'+m_1')(e_2+m_2')(e_3'+m_3')}},
\label{eq:LFformfactor03}
\end{split}
\end{equation}
and
\begin{equation}
\begin{split}
f^V_1(q^2)&=\int\mathcal{DS}_{1}\frac{1}{8\bar{P}^+\bar{P}'^+}
\text{Tr}[(\slashed{\bar{P}}+M_0)\gamma^+(\slashed{\bar{P}}'+M_0')\gamma_{\bot\alpha}\gamma_{5}(\slashed{p}_3'+m_3')\gamma^+(\slashed{p}_3+m_3)\gamma_{\bot\beta}\gamma_{5}],\\
f^V_2(q^2)&=\int\mathcal{DS}_{1}\frac{iM}{8\bar{P}^+\bar{P}'^+q^2}
\text{Tr}[(\slashed{\bar{P}}+M_0)\sigma^{+\mu}q_{\mu}(\slashed{\bar{P}}'+M_0')\gamma_{\bot\alpha}\gamma_{5}(\slashed{p}_3'+m_3')\gamma^+(\slashed{p}_3+m_3)\gamma_{\bot\beta}\gamma_{5}],\\
f^V_3(q^2)&=\frac{M}{M+M'}\left(f_1^V(q^2)(1-\frac{2\bar{P}\cdot q}{q^2})
+\int\frac{\mathcal{DS}_{1}}{4\sqrt{\bar{P}^+\bar{P}'^+}q^2}
\text{Tr}[(\slashed{\bar{P}}+M_0)\gamma^+(\slashed{\bar{P}}'+M_0')\gamma_{\bot\alpha}\gamma_{5}(\slashed{p}_3'+m_3')\slashed{q}\slashed{p}_3+m_3)\gamma_{\bot\beta}\gamma_{5}]\right),\\
g^A_1(q^2)&=\int\mathcal{DS}_{1}\frac{1}{8\bar{P}^+\bar{P}'^+}
\text{Tr}[(\slashed{\bar{P}}+M_0)\gamma^+\gamma_5(\slashed{\bar{P}}'+M_0')\gamma_{\bot\alpha}\gamma_{5}(\slashed{p}_3'+m_3')\gamma^+\gamma_5(\slashed{p}_3+m_3)\gamma_{\bot\beta}\gamma_{5}],\\
g^A_2(q^2)&=\int\mathcal{DS}_{1}\frac{-iM}{8\bar{P}^+\bar{P}'^+q^2}
\text{Tr}[(\slashed{\bar{P}}+M_0)\sigma^{+\mu}q_{\mu}\gamma_5(\slashed{\bar{P}}'+M_0')\gamma_{\bot\alpha}\gamma_{5}(\slashed{p}_3'+m_3')\gamma^+\gamma_5(\slashed{p}_3+m_3)\gamma_{\bot\beta}\gamma_{5}],\\
g^A_3(q^2)&=\frac{M}{M-M'}\left(g^A_1(q^2)(-1+\frac{2\bar{P}\cdot q}{q^2})
+\int\frac{-\mathcal{DS}_{1}}{4\sqrt{\bar{P}^+\bar{P}'^+}q^2}
\text{Tr}[(\slashed{\bar{P}}+M_0)\gamma^+\gamma_5(\slashed{\bar{P}}'+M_0')\gamma_{\bot\alpha}\gamma_{5}(\slashed{p}_3'+m_3')~\slashed{q}~\gamma_5~(\slashed{p}_3+m_3)\gamma_{\bot\beta}\gamma_{5}]\right),\\
\mathcal{DS}_{1}&=\frac{dx_1d^2\vec{k}_{1\bot}dx_2d^2\vec{k}_{2\bot}}{2(2\pi)^32(2\pi)^3}
\frac{\phi^*(x_i',\vec{k}_{i\bot}')\phi(x_i,\vec{k}_{i\bot})}{48\sqrt{x_3x_3'M_0^3M_0'^3}}
\frac{\text{Tr}[\gamma_{\bot}^{\alpha}(\slashed{\bar{P}}'+M_0')(\slashed{p}_1+m_1)(\slashed{\bar{P}}+M_0)\gamma_{\bot}^{\beta}(\slashed{p}_2-m_2)]}{\sqrt{(e_1+m_1)(e_2+m_2)(e_3+m_3)(e_1'+m_1')(e_2+m_2')(e_3'+m_3')}},
\label{eq:LFformfactor04}
\end{split}
\end{equation}
\end{widetext}
for the $\mathcal{B}_b(\bar{3}_f)\to\mathcal{B}_c(\bar{3}_f)$ and $\mathcal{B}_b(6_f)\to\mathcal{B}_c(6_f)$ transitions, respectively.

\section{The semirelativistic potential model for calculating baryon wave function}
\label{sec3}

In this section, we illustrate how to obtain the concerned spatial wave functions by the semirelativistic quark model with the help of the GEM. Different from the meson system, baryon is a typical three-body system. Thus, its wave function can be extracted by solving the three-body Schr\"{o}dinger equation. Here, the semirelativistic potentials were given in Refs. \cite{Godfrey:1985xj,Capstick:1985xss} which are applied to the realistic calculation of this work. The involved Hamiltonian includes \cite{Li:2021qod}
\begin{equation}
\mathcal{H}=K+\sum_{i<j}(S_{ij}+G_{ij}+V^{\text{so(s)}}_{ij}+V^{\text{so(v)}}_{ij}+V^{\text{ten}}_{ij}+V^{\text{con}}_{ij})
\end{equation}
with $K$, $S$, $G$, $V^{\text{so}(s)}$, $V^{\text{so}(v)}$,$V^{\text{tens}}$ and $V^{\text{con}}$ representing the kinetic energy, the spin-independent linear confinement piece, the Coulomb-like potential, the scalar type-spin-orbit interaction, the vector type-spin-orbit interaction, the tensor potential, and the spin-dependent contact potential, respectively. Their concrete expressions are listed here \cite{Godfrey:1985xj,Capstick:1985xss,Song:2015nia,Pang:2017dlw}:
\begin{eqnarray}
K&=&\sum_{i=1,2,3}\sqrt{m_i^2+p_i^2},\\
S_{ij}&=&-\frac{3}{4}\left(br_{ij}\left[\frac{e^{-\sigma^2r_{ij}^2}}{\sqrt{\pi}\sigma r_{ij}}+\left(1+\frac{1}{2\sigma^2r_{ij}^2}\right)\frac{2}{\sqrt{\pi}}\right.\right.\nonumber\\&&\left.\left.
\times\int_{0}^{\sigma r_{ij}}e^{-x^2}dx\right]\right)\mathbf{F_i}\cdot\mathbf{F_j}+\frac{c}{3},\\
G_{ij}&=&\sum_{k}\frac{\alpha_k}{r_{ij}}\left[\frac{2}{\sqrt{\pi}}\int_{0}^{\tau_k r_{ij}}e^{-x^2}dx\right]\mathbf{F_i}\cdot\mathbf{F_j}
\end{eqnarray}
for the spin-independent terms with
\begin{equation} \sigma^2=\sigma_0^2\left[\frac{1}{2}+\frac{1}{2}\left(\frac{4m_im_j}{(m_i+m_j)^2}\right)^4+s^2\left(\frac{2m_im_j}{m_i+m_j}\right)^2\right],
\end{equation}
and
\begin{equation}
\begin{split}
V^{\text{so}(s)}_{ij}=&-\frac{\mathbf{r_{ij}}\times\mathbf{p_{i}}\cdot\mathbf{S_i}}{2m_i^2}\frac{1}{r_{ij}}
\frac{\partial S_{ij}}{r_{ij}}+\frac{\mathbf{r_{ij}}\times\mathbf{p_{j}}\cdot\mathbf{S_j}}{2m_j^2}\frac{1}{r_{ij}}\frac{\partial S_{ij}}{\partial r_{ij}},\\
V^{\text{so}(v)}_{ij}=&\frac{\mathbf{r_{ij}}\times\mathbf{p_{i}}\cdot\mathbf{S_i}}{2m_i^2}\frac{1}{r_{ij}}\frac{\partial G_{ij}}{r_{ij}}
-\frac{\mathbf{r_{ij}}\times\mathbf{p_{j}}\cdot\mathbf{S_j}}{2m_j^2}\frac{1}{r_{ij}}\frac{\partial G_{ij}}{r_{ij}}
\nonumber\\&-\frac{\mathbf{r_{ij}}\times\mathbf{p_{j}}\cdot\mathbf{S_i}-\mathbf{r_{ij}}\times\mathbf{p_{i}}\cdot\mathbf{S_j}}{m_i~m_j}\frac{1}{r_{ij}}\frac{\partial G_{ij}}{\partial r_{ij}},\\
V^{\text{tens}}_{ij}=&-\frac{1}{m_im_j}\left[\left(\mathbf{S_i}\cdot\mathbf{\hat r_{ij}}\right)\left(\mathbf{S_j}\cdot \mathbf{\hat r_{ij}}\right)-\frac{\mathbf{S_i}\cdot\mathbf{S_j}}{3}\right]\left(\frac{\partial^2G_{ij}}{\partial r^2}-\frac{\partial G_{ij}}{r_{ij}\partial r_{ij}}\right),\\
V^{\text{con}}_{ij}=&\frac{2\mathbf{S_i}\cdot\mathbf{S_j}}{3m_i m_j}\nabla^2G_{ij}
\end{split}
\end{equation}
for the spin-dependent terms, where $m_i$ and $m_j$ are the masses of quark $i$ and $j$, respectively. And, we take $\langle\mathbf{F_i}\cdot\mathbf{F_j}\rangle=-2/3$ for quark-quark interaction.

In the following, a general potential which relies on the center-of-mass of interacting quarks and momentum are made up for the loss of relativistic effects in the nonrelativistic limit \cite{Godfrey:1985xj,Capstick:1985xss,Wang:2018rjg,Wang:2019mhs,Duan:2021alw}, that is,
\begin{equation}
\begin{split}
&G_{ij}\to\left(1+\frac{p^2}{E_iE_j}\right)^{1/2} G_{ij}\left(1+\frac{p^2}{E_iE_j}\right)^{1/2},\\
&\frac{V^{k}_{ij}}{m_im_j}\to\left(\frac{m_im_j}{E_iE_j}\right)^{1/2+\epsilon_k}\frac{V^k_{ij}}{m_im_j}\left(\frac{m_im_j}{E_iE_j}\right)^{1/2+\epsilon_k}
\end{split}
\end{equation}
with $E_i=\sqrt{p^2+m_i^2}$, {where subscript $k$ was applied to distinguish the contributions from the contact, tensor, vector spin-orbit, and scalar spin-orbit terms. In addition, $\epsilon_k$ represents the relevant modification parameters, which are collected in Table \ref{parametersofGI}}.

\begin{table}
\centering
\caption{The parameters used in the semirelativistic potential model \cite{Li:2021qod}.}
\label{parametersofGI}
\renewcommand\arraystretch{1.2}
\begin{tabular*}{86mm}{c@{\extracolsep{\fill}}cccc}
\toprule[1pt]
\toprule[0.7pt]
Parameters     & Values   & Parameters   & Values\\
\toprule[0.7pt]
$m_u~(\text{GeV})$     & $0.220$   & $\epsilon^{\text{so}(s)}$   & $0.448$\\
$m_d~(\text{GeV})$     & $0.220$   & $\epsilon^{\text{so}(v)}$   & $-0.062$\\
$m_s~(\text{GeV})$     & $0.419$   & $\epsilon^{\text{tens}}$   & $0.379$\\
$m_c~(\text{GeV})$     & $1.628$   & $\epsilon^{\text{con}}$   & $-0.142$\\
$m_b~(\text{GeV})$     & $4.977$   & $\sigma_0~(\text{GeV})$   & $2.242$\\
$b~(\text{GeV}^2)$     & $0.142$   & $s$   & $0.805$\\
$c~(\text{GeV})  $     & $-0.302$  & $$    & $$\\
\bottomrule[0.7pt]
\bottomrule[1pt]
\end{tabular*}
\end{table}

The total wave function of the single heavy baryon is composed of color, flavor, spatial, and spin wave functions, i.e.,
\begin{equation}
\Psi_{\mathbf{J},\mathbf{M_J}}=\chi^{c}\left\{\chi^{s}_{\mathbf{S},\mathbf{M_S}}\psi^{p}_{\mathbf{L},\mathbf{M_L}} \right\}_{\mathbf{J},\mathbf{M_J}}\psi^{f},
\end{equation}
where $\chi^{c}=(rgb-rbg+gbr-grb+brg-bgr)/\sqrt{6}$ is the color wave function, which is universal for the baryon. For the $\Xi^{(*)}_{Q}$ baryon, its flavor wave function is $\psi^{\text{flavor}}_{\Xi^{(*)}_{Q}}=(ns-sn)Q/\sqrt{2}$, while for the $\Omega_Q^{(*)}$ baryon, its flavor wave function denotes $\psi^{\text{flavor}}_{\Omega^{(*)}_{Q}}=ssQ$, where $Q=b, c$ and $n=u, d$\footnote{
A brief introduction about the classification of the single heavy baryons is helpful to the reader to understand how to construct their wave functions.
The single heavy baryons with one heavy-flavor quark and two light-flavor quarks belong to the symmetric $6_{\text{F}}$ or antisymmetric $\bar{3}_{\text{F}}$ flavor representations based on the flavor SU(3) symmetry. The total color-flavor-spin wave functions for the $S$-wave members must be antisymmetric. Considering the color wave function is antisymmetric invariably, hence the spin of the two light quarks is $S=1$ for $6_{\text{F}}$ (e.g. $\Sigma_Q$, $\Xi_Q^{\prime}$ and $\Omega_Q$) or $S=0$ for $\bar{3}_{\text{F}}$ (e.g. $\Lambda_Q$ and $\Xi_Q$).
More details about the classification of the single heavy baryons can be found in Refs. \cite{Chen:2007xf,Chen:2021eyk}. For $\Xi_Q^{\prime}$, its flavor wave function is  $\psi^{\text{flavor}}_{\Xi^{\prime}_{Q}}=(ns+sn)Q/\sqrt{2}$.
}.
Besides, \textbf{S} denotes the total spin and \textbf{L} is the total orbital angular momentum. $\psi^{p}_{\mathbf{L},\mathbf{M_L}}$ is the spatial wave function which is composed of $\rho$ mode and $\lambda$ mode, that is,
\begin{equation}
\psi^{p}_{\mathbf{L},\mathbf{M_L}}(\vec{\rho},\vec{\lambda})=\left\{\phi_{\pmb{l_{\rho}},\pmb{ml_{\rho}}}(\vec{\rho})
\phi_{\pmb{l_{\lambda}},\pmb{ml_{\lambda}}}(\vec{\lambda})\right\}_{\mathbf{L},\mathbf{M_L}},
\end{equation}
where the subscripts $\pmb{l_{\rho}}$ and $\pmb{l_{\lambda}}$ are the orbital angular momentum quanta for $\rho$ and $\lambda$ mode, respectively, and the internal Jacobi coordinates are chosen as
\begin{equation}
\vec{\rho}=\vec{r}_1-\vec{r}_2,~~\vec{\lambda}=\vec{r}_3-\frac{m_1 \vec{r}_1+m_2 \vec{r}_2}{m_1+m_2}.
\end{equation}
In this work, the Gaussian basis \cite{Hiyama:2003cu,Yoshida:2015tia,Yang:2019lsg},
\begin{equation}
\begin{split}
\phi_{nlm}^{G}(\vec{r})=&\phi^{G}_{nl}(r)~Y_{lm}(\hat{r})\\
=&\sqrt{\frac{2^{l+2}(2\nu_{n})^{l+3/2}}{\sqrt{\pi}(2l+1)!!}}\lim_{\varepsilon\rightarrow 0}\frac{1}{(\nu_{n}\varepsilon)^l}\sum_{k=1}^{k_{\text{max}}}C_{lm,k}e^{-\nu_{n}(\vec{r}-\varepsilon \vec{D}_{lm,k})^2},
\label{Gaussian basis}
\end{split}
\end{equation}
is adopted to expand the spatial wave functions $\phi_{\pmb{l_{\rho}},\pmb{ml_{\rho}}}$ and $\phi_{\pmb{l_{\lambda}},\pmb{ml_{\lambda}}}$ ($n=1,2,\cdots,n_{\rm{max}}$). Here, a freedom parameter $n_{\rm{max}}$ should be chosen from positive integers, and the Gaussian size parameter $\nu_{n}$ is settled as a geometric progression as
\begin{equation}
\nu_{n}=1/r^2_{n}, ~r_{n}=r_{\rm{min}}~a^{n-1}
\end{equation}
with $$a=\left(\frac{r_{max}}{r_{\rm{min}}}\right)^{\frac{1}{n_{\rm{max}}-1}}.$$
Meanwhile, in our calculation the values of $\rho_{\rm{min}}$ and $\rho_{\rm{max}}$ are chosen as $0.2$ and $2.0$~fm, respectively, and $n_{\rho_{\rm{max}}}=6$. For $\lambda$ mode, we also use the same Gaussian sized parameters.

The Rayleigh-Ritz variational principle is used in this work to solve the three-body Schr\"{o}dinger equation
\begin{equation}
\mathcal{H} \Psi_{\mathbf{J},\mathbf{M_J}}=E \Psi_{\mathbf{J},\mathbf{M_J}}.
\end{equation}
Finally, by solving Schr\"{o}dinger equation, the masses and wave functions of the baryons are obtained, which are collected in Table \ref{tab:wavefunctions}.

\begin{table*}[htbp]\centering
\caption{Spatial wave functions of the concerned $\Xi_Q$ and $\Omega_Q$ from the GI model and GEM. It is worth to mention that the masses for the neutral and charged states are degenerate here due to the same masses for $u$ and $d$ quarks. The second column denotes our theoretically prediction, while the third column denotes the experimental data quoted from the PDG \cite{ParticleDataGroup:2020ssz}. Here, the first value in each row is the masses for the neutral baryon, while the second one is the mass for the charged state. The Gaussian bases $(n_{\rho},n_{\lambda})$ listed in the third column are arranged as $[(1,1),(1,2),\cdots,(1,n_{\lambda_{\rm{max}}}),(2,1),(2,2),\cdots,(2,n_{\lambda_{\rm{max}}}),\cdots,(n_{\rho_{\rm{max}}},1),(n_{\rho_{\rm{max}}},2),\cdots,(n_{\rho_{\rm{max}}},n_{\lambda_{\rm{max}}})]$.}
\label{tab:wavefunctions}
\renewcommand\arraystretch{1.2}
\begin{tabular*}{172mm}{c@{\extracolsep{\fill}}cccc}
\toprule[0.5pt]
\toprule[0.5pt]
Baryon  &This work (GeV)  &Experiment (MeV)  &Eigenvector\\
\midrule[0.5pt]
\multirow{4}*{\shortstack{$\Xi_b\left(\frac{1}{2}^{+}\right)$}}        &\multirow{4}*{5.804}    &\multirow{4}*{\makecell[c]{$5791.9\pm0.5$ \\  $5797.0\pm0.6$}}
&$[-0.017, -0.040, -0.075, 0.002, -0.003, 0.001, -0.033, -0.026, -0.004,$\\
&&&$-0.009, 0.004, -0.001, 0.005, -0.266, -0.267, 0.013, -0.009, 0.002,$\\
&&&$0.008, 0.017, -0.363, -0.041, 0.007, -0.001, -0.006, 0.004, -0.023,$\\
&&&$-0.079, 0.014, -0.003, 0.002, 0.001, 0.010, 0.007, -0.003, 0.001]$\\
\multirow{4}*{\shortstack{$\Omega_b\left(\frac{1}{2}^{+}\right)$}}        &\multirow{4}*{6.043}    &\multirow{4}*{\makecell[c]{$6046.1\pm1.7$}}
&$[0.002, 0.004, 0.011, -0.006, 0.003, -0.001, 0.075, -0.024, 0.040,$\\
&&&$0.002, 0.000, -0.000, -0.034, 0.361, 0.096, 0.002, -0.001, 0.001,$\\
&&&$-0.009, -0.022, 0.588, -0.002, 0.011, -0.003, 0.009, -0.025, -0.046,$\\
&&&$0.101, -0.025, 0.006, -0.002, 0.006, 0.008, -0.013, 0.005, -0.001]$\\
\multirow{4}*{\shortstack{$\Xi_c\left(\frac{1}{2}^{+}\right)$}}        &\multirow{4}*{2.474}    &\multirow{4}*{\makecell[c]{$2470.90^{+0.22}_{-0.29}$ \\ \\ $2467.94^{+0.17}_{-0.20}$}}
&$[-0.017, -0.027, -0.082, -0.010, -0.001, 0.000, -0.028, -0.032, -0.010,$\\
&&&$-0.011, 0.004, -0.001, 0.005, -0.192, -0.315, -0.032, -0.000, 0.000,$\\
&&&$0.002, 0.037, -0.297, -0.116, 0.020, -0.005, -0.004, -0.002, -0.010,$\\
&&&$-0.082, 0.010, -0.002, 0.001, 0.002, 0.007, 0.009, -0.003, 0.001]$\\
\multirow{4}*{\shortstack{$\Xi_c^{*}\left(\frac{1}{2}^{+}\right)$}}    &\multirow{4}*{2.947}    &\multirow{4}*{\makecell[c]{$2970.9^{+0.4}_{-0.6}$ \\ \\ $2966.34^{+0.17}_{-1.00}$}}
&$[-0.023, -0.072, -0.098, 0.147, -0.012, 0.003, -0.039, -0.081, -0.007,$\\
&&&$0.048, -0.004, 0.001, 0.015, -0.390, -0.469, 0.501, -0.049, 0.011,$\\
&&&$0.011, 0.013, -0.268, 0.682, -0.023, 0.005, -0.007, -0.005, -0.048,$\\
&&&$0.314, 0.056, -0.010, 0.001, 0.006, 0.012, -0.044, 0.005, -0.000]$\\
\multirow{4}*{\shortstack{$\Omega_c\left(\frac{1}{2}^{+}\right)$}}        &\multirow{4}*{2.692}    &\multirow{4}*{\makecell[c]{$2695.2\pm1.7$}}
&$[0.006, -0.003, 0.019, -0.008, 0.004, -0.001, 0.093, -0.027, 0.045,$\\
&&&$0.001, 0.002, -0.000, -0.049, 0.351, 0.135, 0.029, -0.010, 0.003,$\\
&&&$0.005, -0.078, 0.527, 0.075, -0.002, -0.001, 0.004, -0.001, -0.071,$\\
&&&$0.096, -0.021, 0.005, -0.001, 0.000, 0.013, -0.014, 0.005, -0.001]$\\
\multirow{4}*{\shortstack{$\Omega_c^{*}\left(\frac{1}{2}^{+}\right)$}}    &\multirow{4}*{3.149}    &\multirow{4}*{\makecell[c]{$-$}}
&$[0.022, -0.025, 0.042, -0.016, 0.007, -0.002, 0.100, 0.112, -0.022,$\\
&&&$-0.060, 0.003, -0.000, -0.043, 0.412, 0.494, -0.188, 0.036, -0.008,$\\
&&&$-0.002, 0.032, 0.068, -0.754, 0.052, -0.011, -0.008, 0.019, -0.076,$\\
&&&$-0.375, -0.010, 0.000, 0.003, -0.008, 0.021, 0.036, -0.007, 0.001]$\\
\toprule[0.5pt]
\toprule[0.5pt]
\end{tabular*}
\end{table*}

As collected in the PDG \cite{ParticleDataGroup:2020ssz}, there are ten states in the $\Xi_c$ family, where the ground states includes $\Xi_{c}^{+}$ and $\Xi_{c}^{0}$ with the quark flavor $usc$ and $dsc$, respectively. $\Xi_{c}^{+}$ was first reported by SPEC \cite{Biagi:1983en}, and then confirmed in Ref. \cite{FermilabE687:1992wmm} by analyzing the $\Xi^{-}\pi^{+}\pi^{+}$ final state, while the neutral one $\Xi_{c}^{0}$ was first discovered by CLEO \cite{CLEO:1988yda} in the  $\Xi^{-}\pi^{+}$  mode. The masses fitted by the PDG are $2467.71\pm0.23$ and $2470.44\pm0.28$ MeV for charged $\Xi_{c}^{+}$ and neutral $\Xi_{c}^{0}$, respectively.
And then, the Belle Collaboration found $\Xi_{c}^+(2970)$ and $\Xi_{c}^0(2970)$ in the  $\Lambda_{c}^{+}K^{-}\pi^{+}$ and $\Lambda_{c}^{+}K_S^{0}\pi^{-}$ final states \cite{Belle:2006edu}, respectively, where the masses of the charged and neutral $\Xi_c(2970)$ states are measured to be  $2964.3\pm1.5$ and $2967.1\pm1.7$ MeV, respectively. As indicated by our calculation shown in Table \ref{tab:wavefunctions}, the observed $\Xi_{c}(2970)$ are good candidate of $\Xi_c^{*}(2S)$.
The ground $\Omega_c$ state, denoted as $\Omega_c(\frac{1}{2}^+)$, was firstly observed in the $\Xi^-K^-\pi^+\pi^+$ channel by WA62 \cite{Biagi:1984mu}, and then was confirmed in ARGUS \cite{ARGUS:1992mwl} by checking the same mode. Its mass was fitted as $2695.2\pm1.7$ MeV by the PDG. Our result given in Table \ref{tab:wavefunctions} indeed supports this assignment since the calculated mass of $\Omega_c(\frac{1}{2}^+)$ is 2.692 GeV consistent with the experimental data.
For the $\Omega_c^*(\frac{1}{2}^+)$ state, which is the first radial excitation of $\Omega_c(\frac{1}{2}^+)$,
its mass is calculated to be 3.149 GeV\footnote{In 2017, the LHCb Collaboration \cite{LHCb:2017uwr} announced that five narrow excited $\Omega_c$ states, $\Omega_c(3000)$, $\Omega_c(3050)$, $\Omega_c(3066)$, $\Omega_c(3090)$, and $\Omega_c(3119)$, were found in the  $\Xi_c^-K^+$ invariant mass spectrum. Later, Belle \cite{Belle:2017ext} confirmed four narrow excited $\Omega_c$ states in the same mode. The spin-parity of these excited strange charmed baryons are not measured yet.
In these five excited $\Omega_c$ states, the masses of $\Omega_c(3090)$ and $\Omega_c(3119)$ were measured as $3090.0\pm0.5$ and $3119.1\pm1.0$ MeV, respectively. Their structures were discussed by various theoretical approaches \cite{Chen:2017gnu,Cheng:2017ove,Chen:2017sci,Wang:2017hej,Agaev:2017jyt,Debastiani:2018adr}. Chen {\it et al.} \cite{Chen:2017gnu} indicated that $\Omega_c(3119)$ cannot be a $2S$ candidate by performing an analysis of the mass spectrum and decay behavior. Cheng {\it et al.} \cite{Cheng:2017ove} assigned $\Omega_c(3090)$ and $\Omega_c(3119)$ as the first radially excited states with $J^P=1/2^+$ and $3/2^+$, respectively, by the analysis of the Regge trajectories and
a direct calculation of the mass via a quark-diquark model. Wang {\it et al.} \cite{Wang:2017hej} proposed that the $\Omega_c(3119)$ favors the $2S$ assignment by a study with a constituent quark model.
Agaev {\it et al.} \cite{Agaev:2017jyt} discussed the favored assignment $\Omega_c(2S)$ state with $J^P=1/2^+$ and $3/2^+$ for $\Omega_c(3066)$ and $\Omega_c(3119)$ with QCD sum rules. Thus, establishing $\Omega_c^*(\frac{1}{2}^+)$ state is still ongoing. In this work, we adopt the calculated result as mass input of the $\Omega_c^*(\frac{1}{2}^+)$ state.}.

In Table \ref{tab:wavefunctions}, we also collected the numerical spatial wave functions corresponding to these charmed baryons, which will be applied to the
following study.

\section{The form factors and color-allowed two-body nonleptonic decays}
\label{sec4}

\subsection{The weak transitions form factors}

With the input of these obtained numerical wave functions of bottom (see Table \ref{tab:wavefunctions}) and charmed baryons, and the expressions of the form factors [see Eqs. (\ref{eq:LFformfactor03})-(\ref{eq:LFformfactor04})], we present the numerical results for the weak transition form factors of $\Xi_{b}\to\Xi_{c}^{(*)}(1/2^+)$ and $\Omega_{b}\to\Omega_{c}^{(*)}(1/2^+)$ processes. {Since the expressions of form factors in Eqs. (\ref{eq:LFformfactor01})-(\ref{eq:LFformfactor04}) are working in the spacelike region ($q^2<0$), we need to extend them to the timelike region ($q^2>0$).} The dipole form \cite{Li:2021qod,Ke:2019smy,Chua:2018lfa,Chua:2019yqh}
\begin{equation}
F(q^2)=\frac{F(0)}{(1-q^2/M^2)[1-b_{1}(q^2/M^2)+b_{2}(q^2/M^2)^2]}
\label{eq:formfactor}
\end{equation}
is applied in this work, {where $F(0)$ is the form factor at $q^2=0$, 
$b_1$, and $b_2$ are obtained by computing each form factor by Eqs.
(\ref{eq:LFformfactor03})-(\ref{eq:LFformfactor04}) from $q^2=-q_{\text{max}}^2$  to $q^2=0$, and fit them by Eq. (\ref{eq:formfactor}) .

With the spatial wave functions obtained in the last subsection, we can calculate out the form factors numerically in the framework of the three-body light-front quark model. In this way, all free parameters of the semirelativistic potential model can be fixed by reproducing the mass spectrum of observed heavy baryons. In the previous work  \cite{Guo:2005qa,Zhu:2018jet,Zhao:2018zcb,Chua:2018lfa,Chua:2019yqh,Ke:2019smy} on baryon weak transitions, simple hadronic oscillator wave function with the oscillator parameter $\beta$ was widely used to simulate the baryon spatial wave function. This treatment makes the results dependent on $\beta$ value. In this work, our study is supported by hadron spectroscopy. Thus, we can avoid the above uncertainty resulted by the selection of spatial wave functions of heavy baryons involved in these discussed transitions.

The extended form factors of $\Xi_b\to\Xi_c^{(*)}$ are collected in Table \ref{tab:XiQformfactors}. The $q^2$ dependence of $f^V_{1,2,3}$ and $g^A_{1,2,3}$ for the $\Xi_{b}\to\Xi_{c}$ and $\Xi_{b}\to\Xi_{c}(2970)$ transitions are plotted in Fig. \ref{fig:Xib2Xicformfactors}.

\begin{table}[htbp]\centering
\caption{{The form factors for the $\Xi_b\to\Xi_c^{(*)}$ transitions in the standard light front quark model. Here, we adopt the form defined in Eq. (\ref{eq:formfactor}) for analyzing these form factors.}}
\label{tab:XiQformfactors}
\renewcommand\arraystretch{1.2}
\begin{tabular*}{86mm}{c@{\extracolsep{\fill}}cccc}
\toprule[0.5pt]
\toprule[0.5pt]
        &$F(0)$&$F(q^2_{\text{max}})$&$b_1$&$b_2$\\
\midrule[0.5pt]
&\multicolumn{4}{c}{$\Xi_b\rightarrow\Xi_c$}\\
$f^V_1$  &$0.481$   &$1.015$   &$0.970$   &$0.233$\\
$f^V_2$  &$-0.127$  &$-0.312$  &$1.380$   &$0.578$\\
$f^V_3$  &$-0.046$  &$-0.097$  &$1.187$   &$0.875$\\
$g^A_1$  &$0.471$   &$0.978$   &$0.929$   &$0.226$\\
$g^A_2$  &$-0.026$  &$-0.068$  &$1.318$   &$0.122$\\
$g^A_3$  &$-0.154$  &$-0.377$  &$1.493$   &$0.947$\\
&\multicolumn{4}{c}{$\Xi_b\rightarrow\Xi_c(2970)$}\\
$f^V_1$  &$0.214$   &$0.200$   &$-1.146$  &$2.282$\\
$f^V_2$  &$-0.072$  &$-0.081$  &$-0.356$  &$1.600$\\
$f^V_3$  &$-0.111$  &$-0.221$  &$1.444$   &$0.168$\\
$g^A_1$  &$0.204$   &$0.186$   &$-1.269$  &$2.474$\\
$g^A_2$  &$-0.087$  &$-0.231$  &$1.867$   &$-0.907$\\
$g^A_3$  &$-0.095$  &$-0.113$  &$-0.022$   &$1.687$\\
\bottomrule[0.5pt]
\bottomrule[0.5pt]
\end{tabular*}
\end{table}

\begin{figure*}[htbp]\centering
  \begin{tabular}{cc}
  \includegraphics[width=76mm]{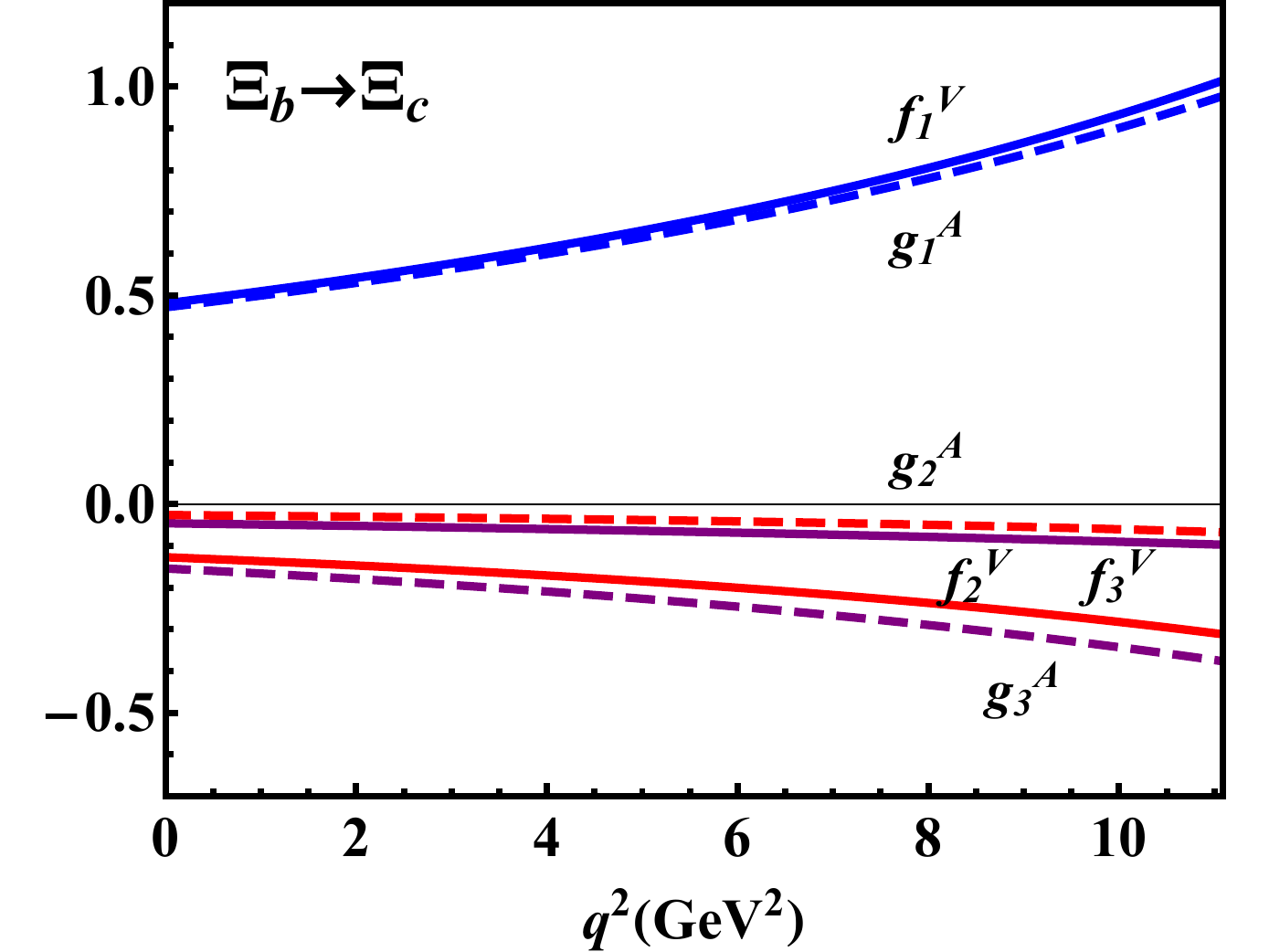}
  \includegraphics[width=76mm]{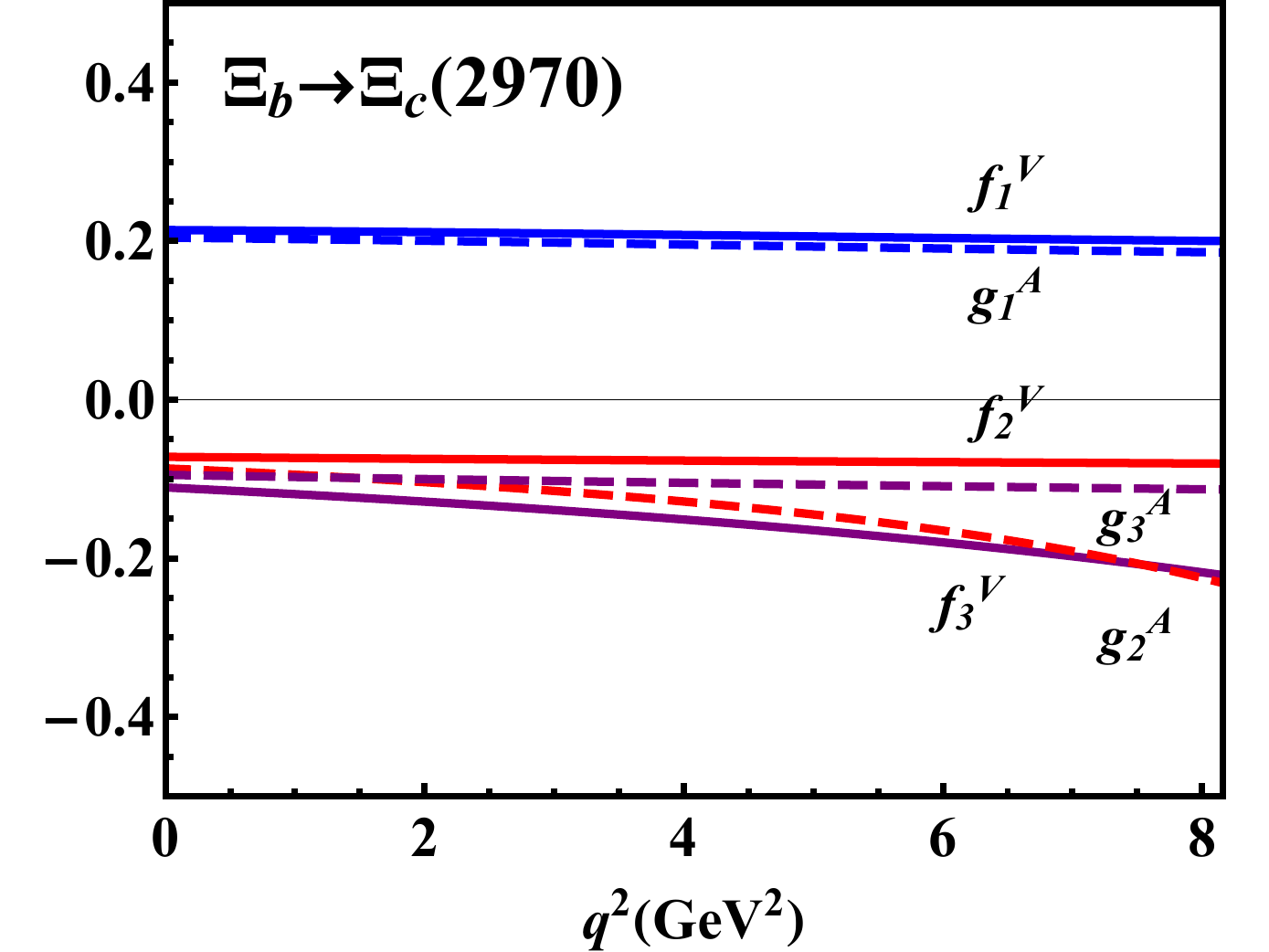}\\
  \end{tabular}
  \caption{The $q^2$ dependence of the form factors $f_{1,2,3}^{V}(q^2)$ and $g_{1,2,3}^{A}(q^2)$ for the $\Xi_{b}\to\Xi_{c}$ (left) and $\Xi_{b}\to\Xi_{c}(2970)$ (right) transitions. Here, the solid and dashed lines represent the vector-type and pseudoscalar-type form factors denoting by the subscripts $V$ and $A$, respectively, while the blue, red, and purple lines (both solid and dashed lines) represent the $i$th form factors denoting by the subscripts respectively for each types.}
\label{fig:Xib2Xicformfactors}
\end{figure*}

For the $\Xi_{b}\to\Xi_{c}$ transition, the corresponding transition matrix element can be rewritten as \cite{Georgi:1990ei,Bowler:1997ej,Chua:2019yqh}
\begin{equation}
\left<\Xi_c(1/2^{+})(\nu')|\bar{c}_{\nu'}\Gamma b_{\nu}|\Xi_b(1/2^{+})(\nu)\right>=\zeta(\omega)\bar{u}(\nu')\Gamma u(\nu),
\end{equation}
in the heavy quark limit at the leading order, so the form factors have more simple behaviors as
\begin{equation}
\begin{split}
&f^V_1(q^2)=g^A_1(q^2)=\zeta(\omega),\\
&f^V_2=f^V_3=g^A_2=g^A_3=0,
\end{split}
\end{equation}
where $\omega=\nu\cdot\nu'=(M^2+M'^2-q^2)/(2MM')$ with $\nu'=p'/M'$ and $\nu=p/M$ denoting the four velocities for $\Xi_c$ and $\Xi_b$, respectively. $\zeta(\omega)$ is the well-known Isgur-Wise function (IWF) and usually expressed as a Taylor series expansion as
\begin{equation}
\zeta(\omega)=1-\zeta_{1}(\omega-1)+\frac{\zeta_{2}}{2}(\omega-1)^2+\cdots,
\label{eq:IWFunction}
\end{equation}
where $\zeta_{1}=-\frac{d\zeta(\omega)}{d\omega}|_{\omega=1}$ and $\zeta_{2}=\frac{d^2\zeta(\omega)}{d\omega^2}|_{\omega=1}$ are two shape parameters depicting the IWF. The most obvious character is in the point $q^2=q_{\rm{max}}^2=(M-M')^2$ (or $\omega=1$),
\begin{equation*}
f^V_1(q_{\rm{max}}^2)=g^A_1(q_{\rm{max}}^2)=\zeta(1)=1.
\end{equation*}
It provided one strong restriction for our result. Besides, when comparing our results with the predictions in heavy quark limit (HQL), we can conclude that our results can well match the requirement from heavy quark effective theory, i.e.,
\begin{enumerate}
\item[{\bf 1.}] $f^V_1$ and $g^A_1$ are close to each other, and dominate over $f^V_{2,3}$ and $g^A_{2,3}$.
\item[{\bf 2.}] At $q^2=q_{\rm{max}}^2$, $f^V_1(q_{\rm{max}}^2)=1.015$ and $g^A_1(q_{\rm{max}}^2)=0.978$ are very approach to 1.
\end{enumerate}
In addition we also extract the two IWF's shape parameters $\xi_1$ and $\xi_2$ in Eq. (\ref{eq:IWFunction}) by fitting $\zeta(\omega)$ from $f^V_1(q^2)$ and $g^A_1(q^2)$, respectively. The concrete results and other theoretical predictions are listed in Table \ref{tab:IWfunction}.

\begin{table}[htbp]\centering
\caption{Our results for the IWF's shape parameters of the $\Xi_b\to\Xi_c$ transition. The superscripts $[a]$ and $[b]$ in the second and third rows represent the fitting of $f_1^{V}$ and $g_1^{A}$, respectively.}
\label{tab:IWfunction}
\renewcommand\arraystretch{1.2}
\begin{tabular*}{86mm}{c@{\extracolsep{\fill}}ccc}
\toprule[0.5pt]
\toprule[0.5pt]
                     &$\zeta_{1}$   &$\zeta_{2}$ \\
\midrule[0.5pt]
This work$^{[a]}$    &1.97          &3.28\\
This work$^{[b]}$    &2.23          &4.63\\
RQM \cite{Ebert:2006rp}  &2.27      &7.74\\
\bottomrule[0.5pt]
\bottomrule[0.5pt]
\end{tabular*}
\end{table}

For the $\Xi_b\to\Xi_c(2970)$ transition, the HQL requires $f_1^V=g_1^A=0$ at $q^2=q_{\text{max}}^2$ since the wave functions of the low-lying $\Xi_b$ and the radial excited state $\Xi_c^{*}(2S)$ are orthogonal \cite{Chua:2019yqh}. Evidently, our results well embody this prediction according to Fig. \ref{fig:Xib2Xicformfactors}.

\begin{table}[htbp]\centering
\caption{The form factors for the $\Omega_b\to\Omega_c^{(*)}$ transitions in the standard light front quark model. We use a three parameter form defined in Eq. (\ref{eq:formfactor}) for these form factors.}
\label{tab:OmegaQformfactors}
\renewcommand\arraystretch{1.2}
\begin{tabular*}{86mm}{c@{\extracolsep{\fill}}cccc}
\toprule[0.5pt]
\toprule[0.5pt]
        &$F(0)$&$F(q^2_{\text{max}})$&$b_1$&$b_2$\\
\midrule[0.5pt]
&\multicolumn{4}{c}{$\Omega_b\to\Omega_c$}\\
$f^V_1$  &$0.493$   &$1.232$   &$1.765$   &$1.272$\\
$f^V_2$  &$0.436$   &$1.075$   &$1.658$   &$1.001$\\
$f^V_3$  &$-0.255$  &$-0.620$  &$1.628$   &$1.005$\\
$g^A_1$  &$-0.161$  &$-0.329$  &$1.053$   &$0.337$\\
$g^A_2$  &$0.011$   &$0.018$   &$0.822$   &$1.526$\\
$g^A_3$  &$0.055$   &$0.137$   &$1.680$   &$1.052$\\
&\multicolumn{4}{c}{$\Omega_b\to\Omega_c(2S)$}\\
$f^V_1$  &$0.180$   &$0.163$   &$-1.135$  &$3.320$\\
$f^V_2$  &$0.133$   &$0.107$   &$-1.727$  &$4.270$\\
$f^V_3$  &$-0.150$  &$-0.215$  &$0.481$   &$0.239$\\
$g^A_1$  &$-0.058$  &$-0.047$  &$-1.701$  &$3.487$\\
$g^A_2$  &$0.029$   &$0.053$   &$1.455$   &$0.772$\\
$g^A_3$  &$0.023$   &$0.023$   &$-0.671$  &$2.407$\\
\bottomrule[0.5pt]
\bottomrule[0.5pt]
\end{tabular*}
\end{table}

\begin{figure*}[htbp]\centering
  \begin{tabular}{cc}
  \includegraphics[width=76mm]{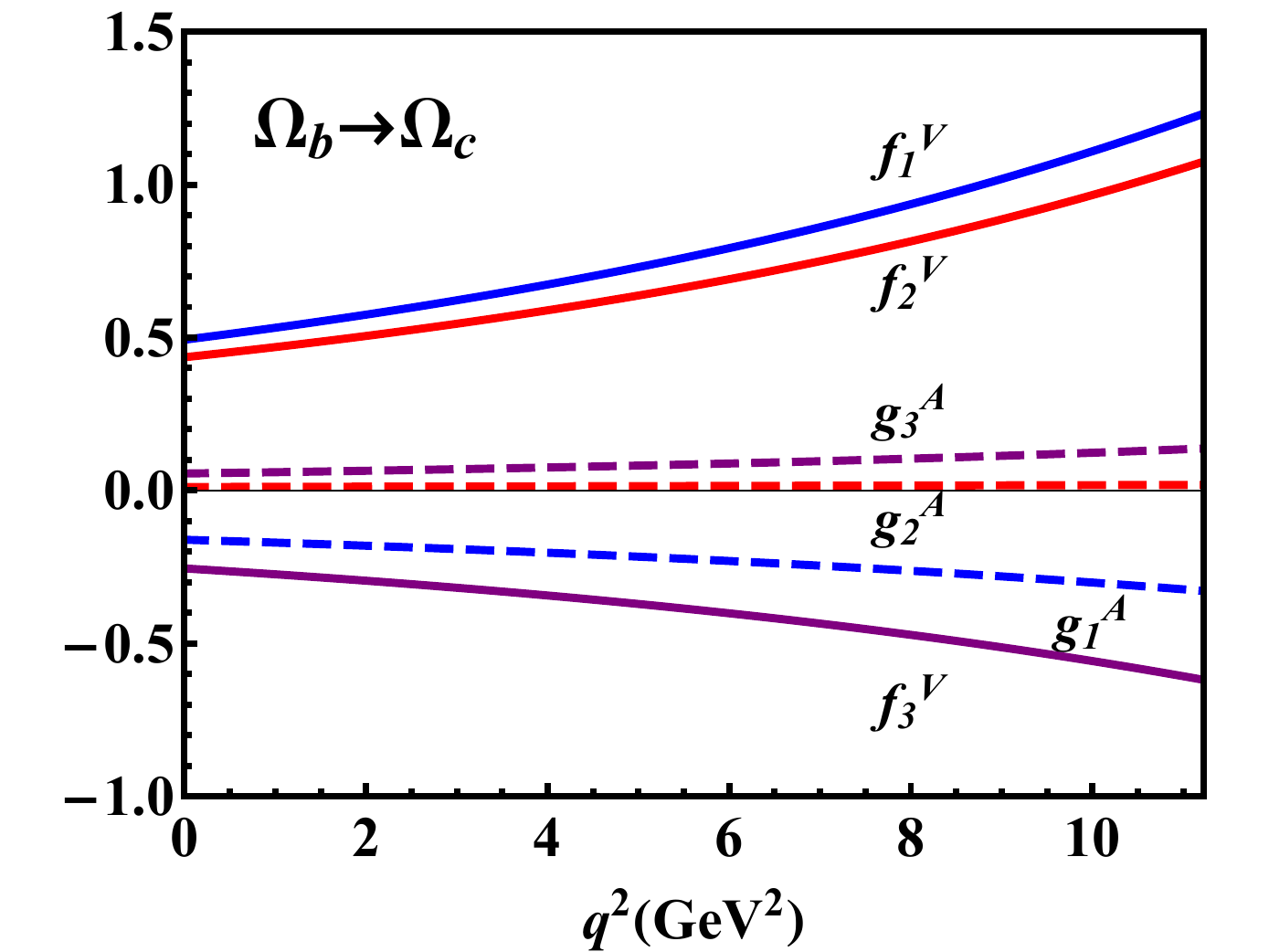}
  \includegraphics[width=76mm]{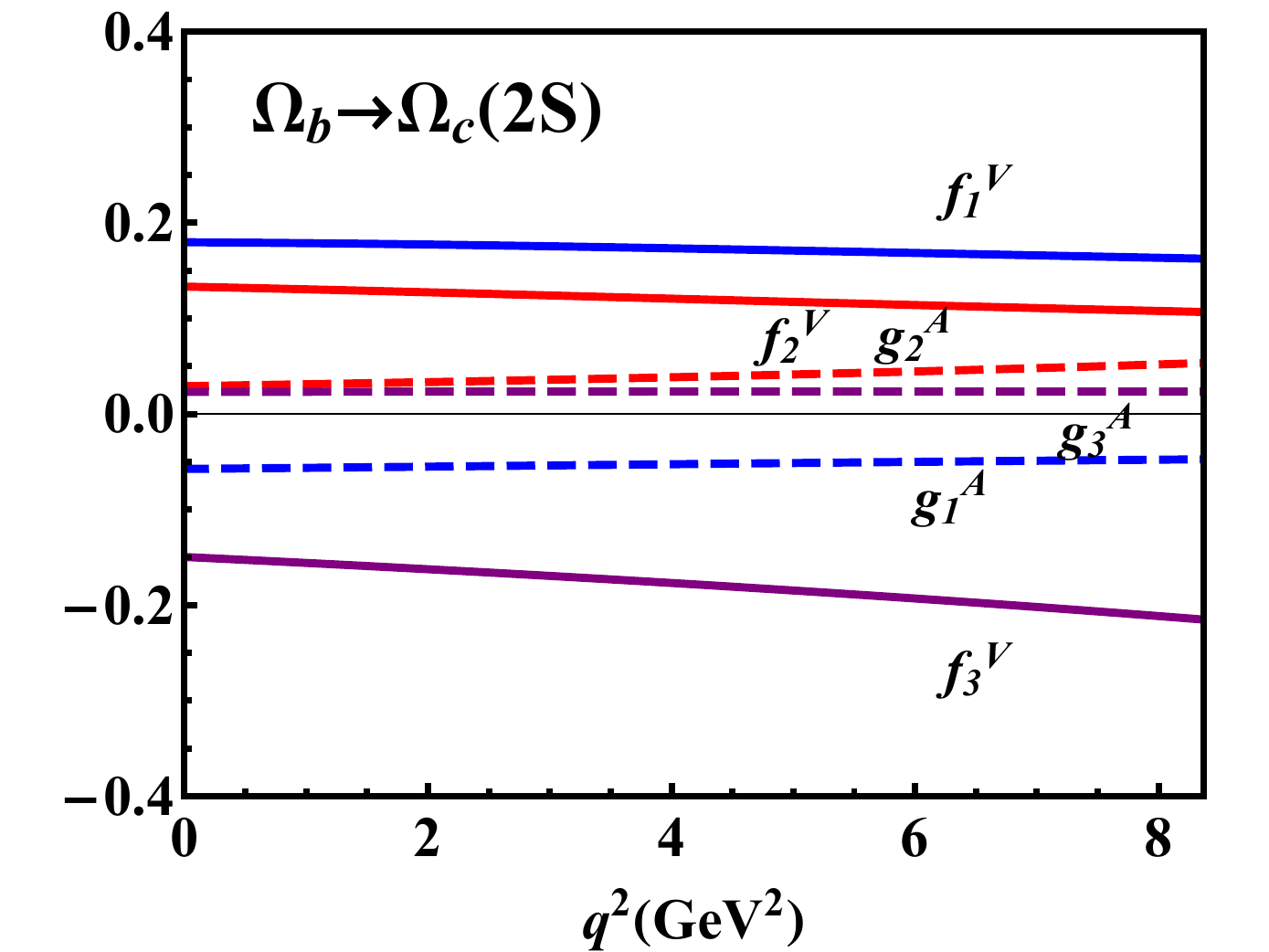}\\
  \end{tabular}
  \caption{The $q^2$ dependence of the form factors $f_{1,2,3}^{V}(q^2)$ and $g_{1,2,3}^{A}(q^2)$ for $\Omega_{b}\to\Omega_{c}$ (left) and $\Omega_{b}\to\Omega_{c}(2S)$ (right) transitions, in which the solid and dashed lines represent the vector or pseudoscalar-types form factors denoting by the subscripts $V$ and $A$, respectively, while the blue, red and purple lines (both solid and dashed lines) represent the $i$th form factors denoting by the subscripts, respectively, for each types.}
\label{fig:Omegab2Omegacformfactors}
\end{figure*}

Additionally, the extended form factors of $\Omega_b\to\Omega_c^{(*)}$ are collected in Table \ref{tab:OmegaQformfactors}. The $q^2$ dependence of $f^V_{1,2,3}$ and $g^A_{1,2,3}$ for the $\Omega_{b}\to\Omega_{c}$ and $\Omega_{b}\to\Omega_{c}(3090)$ transitions are plotted in Fig. \ref{fig:Omegab2Omegacformfactors}. For the $\Omega_{b}\to\Omega_{c}$ transition, the corresponding transition matrix element can be rewritten as \cite{Georgi:1990ei,Bowler:1997ej,Chua:2019yqh}
\begin{equation}
\begin{split}
\langle\Omega_c&(1/2^{+})(\nu')|\bar{c}_{\nu'}\Gamma b_{\nu}|\Omega_b(1/2^{+})(\nu)\rangle=\\
&-\frac{1}{3}(g^{\rho\sigma}\xi_1-v^{\rho}v^{\prime\sigma}\xi_2)\bar{u}(v^{\prime})(\gamma_{\rho}-v^{\prime}_{\rho})\Gamma(\gamma_{\sigma}-v_{\sigma})u(v)
\end{split}
\end{equation}
in HQL at the leading order. Thus, the form factors in HQL have more simple behaviors as
\begin{equation}
\begin{split}
&f_1^V(q_{\text{max}}^2)=\frac{1}{3}+\frac{1}{3}\frac{M^2+M^{\prime2}}{MM^{\prime}}=1.23,\\
&f_2^V(q_{\text{max}}^2)=\frac{1}{3}\frac{M+M^{\prime}}{M^{\prime}}=1.08,\\
&f_3^V(q_{\text{max}}^2)=-\frac{1}{3}\frac{M-M^{\prime}}{M^{\prime}}=-0.41,\\
&g_1^A(q_{\text{max}}^2)=-\frac{1}{3},\\
&g_2^A(q_{\text{max}}^2)=g_3^A(q_{\text{max}}^2)=0,
\label{eq:HQL}
\end{split}
\end{equation}
at $q^2=q_{\text{max}}^2$ point by substituting the involved masses. Obviously, our results located in the third column of the Table \ref{tab:OmegaQformfactors} match well the requirement from the HQL as shown in Eq. (\ref{eq:HQL}), which can be as a direct test to the HQL.

\subsection{The color-allowed two-body nonleptonic decays}

With the preparation of the obtained form factors, we further calculate the color-allowed two-body nonleptonic decays of $\Xi_b$ and $\Omega$ with emitting a pseudoscalar meson ($\pi^{-}$, $K^{-}$, $D^{-}$, and $D_s^{-}$) or a vector meson ($\rho^{-}$, $K^{*-}$, $D^{*-}$, and $D_s^{*-}$). In this work, the decay rates are investigated by the na\"{i}ve factorization approach\footnote{The na\"{i}ve factorization approach works well for the color-allowed dominated processes. But, there exists the case that the color-suppressed and penguin dominated processes can not be explained by the na\"{i}ve factorization, which may show important nonfactorizable contributions to nonleptonic decays \cite{Zhu:2018jet}. As indicated in Refs. \cite{Lu:2009cm,Chua:2018lfa,Chua:2019yqh}, the nonfactorizable contributions in bottom baryon nonleptonic decays are cosiderable comparing with the factorized ones. Since a precise study of nonfactorizable contributions is beyond the scope of the present work, we still adopt the na\"{i}ve factorization approximation.}.

Generally, in the na\"{i}ve factorization assumption, the hadronic transition matrix element is factorized into a product of two independent matrix elements \cite{Ke:2019smy}
\begin{equation}
\begin{split}
\langle\mathcal{B}_c^{(*)}&(P^{\prime},J_{z}^{\prime}) \ M^{-} \ |\mathcal{H}_{\text{eff}}| \ \mathcal{B}_b(P,J_{z})\rangle\\
=&\frac{G_F}{\sqrt{2}} V_{cb} V_{qq^{\prime}}^{*}
\langle M^{-}|\bar{q}^{\prime}\gamma_{\mu}(1-\gamma_{5})q|0\rangle\\
&\times\langle\mathcal{B}_c^{(*)}(P^{\prime},J_{z}^{\prime})|\bar{c}\gamma^{\mu}(1-\gamma_{5})b|\mathcal{B}_b(P,J_{z})\rangle,
\end{split}
\end{equation}
where the meson transition term is given by
\begin{equation}
\begin{split}
\langle M|\bar{q}^{\prime}&\gamma_{\mu}(1-\gamma_{5})q|0\rangle
=\left\{
\begin{array}{ll}
if_{P}q_{\mu}, &M=P\\
if_{V}\epsilon_{\mu}^{*}m_{V}, &M=V
\end{array}.
\right.
\end{split}
\end{equation}
Here, $P$ and $V$ denote pseudoscalar and vector mesons, respectively. The baryon transition term can be obtained by Eq. (\ref{eq:formfactors}). The corresponding Feynman diagram (taking the $\Xi_b^-\to\Xi_c^0M^-$ as an example here) is displayed in Fig. \ref{fig:nonlepton}.

\begin{figure}[htbp]\centering
  \includegraphics[width=6cm]{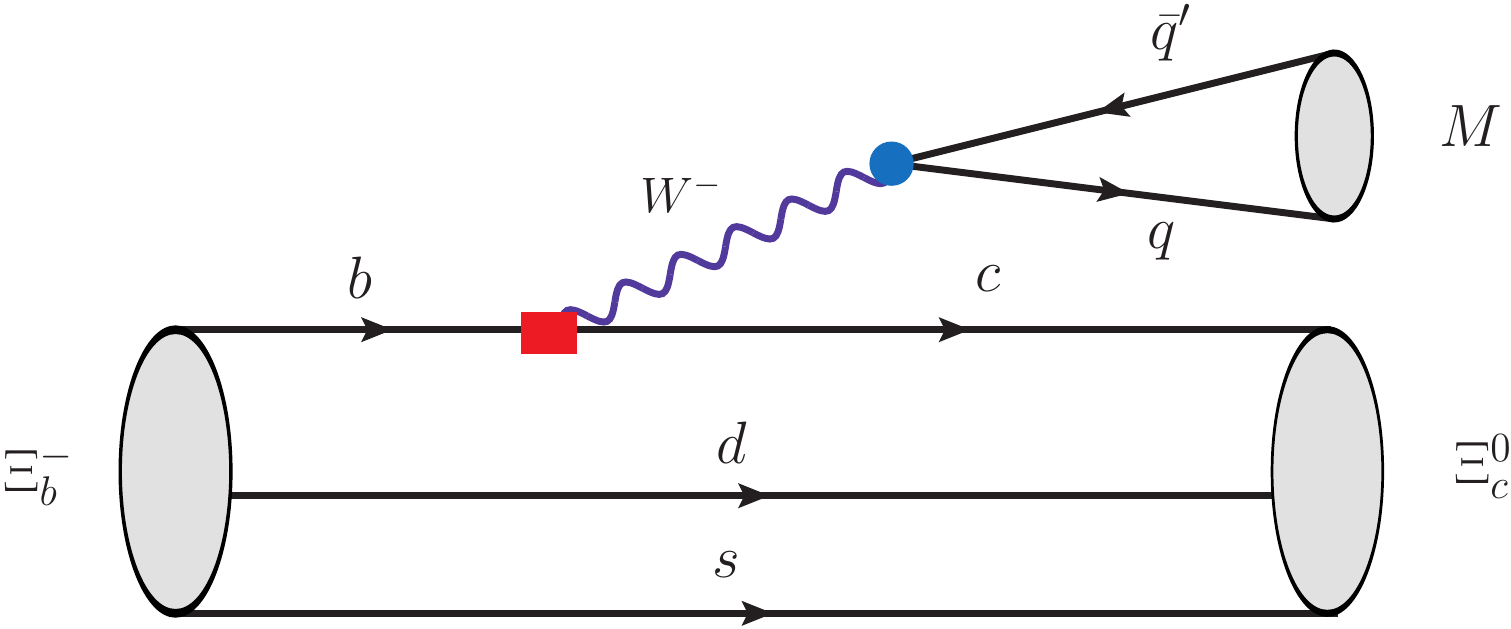}\\
  \caption{The diagram for depicting the color-allowed two-body nonleptonic decay $\Xi_b^-\to\Xi_c^0M^-$ in the tree level.}
  \label{fig:nonlepton}
\end{figure}

Finally the decay width and asymmetry parameter are given by \cite{Ke:2019smy}
\begin{equation}
\begin{split}
\Gamma&=\frac{|p_c|}{8\pi}\left(\frac{(M+M^{\prime})^2-m^2}{M^2}|A|^2+\frac{(M-M^{\prime})^2-m^2}{M^2}|B|^2\right),\\
\alpha&=\frac{2\kappa\text{Re}(A^*B)}{|A|^2+\kappa^2|B|^2},
\label{eq:PhysicsP}
\end{split}
\end{equation}
\begin{equation}
\begin{split}
\Gamma&=\frac{|p_c|(E^{\prime}+M^{\prime})}{4\pi M}\left(2(|S|^2+|P_2|^2)+\frac{E_m^2}{m^2}(|S+D|^2+|P_1|^2)\right),\\
\alpha&=\frac{4m^2\text{Re}(S^*P_2)+2E_m^2\text{Re}(S+D)^*P_1}{2m^2(|S|^2+|P_2|^2)+E_m^2(|S+D|^2+|P_1|^2)},
\label{eq:PhysicsV}
\end{split}
\end{equation}
for the cases involved in the pseudoscalar and vector meson final state, respectively, where $p_c$ is the momentum of the daughter baryon in the rest frame of the parent baryon and $\kappa=|p_c|/(E^{\prime}+M^{\prime})$. Besides, $M(E)$ and $M^{\prime}(E^{\prime})$ are the masses (energies) of the parent (daughter) baryons, respectively, while $m(E_m)$ denotes the mass(energy) of the meson in the final state.

$A$ and $B$ in Eqs. (\ref{eq:PhysicsP}) are given by
\begin{equation}
\begin{split}
A&=\frac{G_F}{\sqrt{2}}V_{cb}V_{qq^{\prime}}^*a_{i}f_{P}(M-M^{\prime})f_1^V(m^2),\\
B&=-\frac{G_F}{\sqrt{2}}V_{cb}V_{qq^{\prime}}^*a_{i}f_{P}(M+M^{\prime})g_1^A(m^2),
\end{split}
\end{equation}
and $S$, $P_{1,2}$ and $D$ in Eqs. (\ref{eq:PhysicsV}) are expressed as
\begin{equation}
\begin{split}
S&=A_1,\\
P_1&=-\frac{|p_c|}{E_m}\left(\frac{M+M^{\prime}}{E^{\prime}+M^{\prime}}B_1+MB_2\right),\\
P_2&=\frac{|p_c|}{E^{\prime}+M^{\prime}}B_1,\\
D&=\frac{|p_c|^2}{E_m(E^{\prime}+M^{\prime})}(A_1-MA_2)
\end{split}
\end{equation}
with
\begin{equation}
\begin{split}
A_{1}&=\frac{G_F}{\sqrt{2}}V_{cb}V_{qq^{\prime}}^*a_{i}f_{V}m_{V}\left(g_1^A(m^2)+g_2^A(m^2)\frac{M-M^{\prime}}{M}\right),\\
A_{2}&=\frac{G_F}{\sqrt{2}}V_{cb}V_{qq^{\prime}}^*a_{i}f_{V}m_{V}\left(2g_{2}^A(m^2)\right),\\
B_{1}&=\frac{G_F}{\sqrt{2}}V_{cb}V_{qq^{\prime}}^*a_{i}f_{V}m_{V}\left(f_1^V(m^2)-f_2^V(m^2)\frac{M+M^{\prime}}{M}\right),\\
B_{2}&=\frac{G_F}{\sqrt{2}}V_{cb}V_{qq^{\prime}}^*a_{i}f_{V}m_{V}\left(2f_{2}^V(m^2)\right),
\end{split}
\end{equation}
where $a_{1}=c_1+c_2/N\approx1.018$ and $a_{2}=c_2+c_1/N\approx0.170$ \cite{Chua:2019yqh}.

With the na\"{i}ve factorization, the color-allowed two-body nonleptonic decays by emitting one pseudoscalar meson or vector meson are presented.
The lifetimes of $\Xi_{b}^{-,0}$ and $\Omega_{b}^{-}$ was reported by the LHCb \cite{LHCb:2014chk,LHCb:2014wqn,LHCb:2016coe} and CDF  \cite{CDF:2014mon} collaborations. In this work, we use the central values as
\begin{equation*}
\tau_{\Xi_b^{0}}=1.480~\rm{fs}, ~~\tau_{\Xi_b^{-}}=1.572~\rm{fs},~~\tau_{\Omega_b^{-}}=1.65~\rm{fs},
\end{equation*}
averaged by the PDG \cite{ParticleDataGroup:2020ssz}. Besides, the masses of the concerned baryons are from the GEM calculation and the Cabibbo-Kobayashi-Maskawa matrix elements
\begin{equation*}
\begin{split}
& V_{cb}=0.0405,\ V_{ud}=0.9740,\ V_{us}=0.2265,\\
& V_{cd}=0.2264,\ V_{cs}=0.9732,
\end{split}
\end{equation*}
are taken from the PDG \cite{ParticleDataGroup:2020ssz}. The decay constants of pseudoscalar and vector mesons include \cite{Cheng:2003sm,Chua:2019yqh}
\begin{equation*}
\begin{split}
&f_{\pi}=130.2,\ f_{K}=155.6,\ f_{D}=211.9,\ f_{D_s}=249.0,\\
&f_{\rho}=216,\ f_{K^*}=210,\ f_{D^*}=220,\ f_{D_s^*}=230,
\end{split}
\end{equation*}
in the unit of MeV.

By substituting our numerical results of the form factors from the three-body light-front quark model and the presented decay parameters into Eqs. (\ref{eq:PhysicsP})-(\ref{eq:PhysicsV}), the branching ratios and asymmetry parameters can be further obtained, which are collected in Tables \ref{tab:nonleptonXi}-\ref{tab:nonleptonOmega} for the $\Xi_b\to\Xi_c^{(*)}$ and $\Omega_b\to\Omega_c^{(*)}$ transitions with emitting a pseudoscalar meson ($\pi^{-}$, $K^{-}$, $D^{-}$, and $D_s^{-}$) or a vector meson ($\rho^{-}$, $K^{*-}$, $D^{*-}$, and $D_s^{*-}$), respectively.

In Table \ref{tab:Comparison}, we compare our results of $\mathcal{B}(\Xi_b^{0,-}\to\Xi_c^{+,0}M^-)$ and $\mathcal{B}(\Omega_b^-\to\Omega_c^0M^-)$ with other theoretical results from the nonrelativistic quark model \cite{Cheng:1996cs}, the relativistic three-quark model \cite{Ivanov:1997hi,Ivanov:1997ra}, the light-front quark model  \cite{Zhao:2018zcb,Chua:2019yqh}, and the covariant confined quark model \cite{Gutsche:2018utw}. Our results are comparable with those calculated from other approaches.
We also notice that the concerned transitions with emitting $\pi^-$, $\rho^-$, and $D_s^{(*)-}$ meson have considerable widths, which are worthy to be explored in future experiment like LHCb and Belle II.

\begin{table*}[htbp]\centering
\caption{The branching ratios and asymmetry parameters of the $\Xi_b\to\Xi_c^{(*)}M$ transitions with $M$ denoting a pseudoscalar or vector meson, where the branching ratios out of or in brackets correspond to the $\Xi_{b}^{0}\rightarrow\Xi_{c}^{+}$ and $\Xi_{b}^{-}\rightarrow\Xi_{c}^{0}$ transitions, respectively.}
\label{tab:nonleptonXi}
\renewcommand\arraystretch{1.2}
\begin{tabular*}{172mm}{l@{\extracolsep{\fill}}ccl@{\extracolsep{\fill}}cc}
\toprule[0.5pt]
\toprule[0.5pt]
Mode                                     &$\mathcal{B}\ (\times10^{-3})$   &$\alpha$
&Mode                                    &$\mathcal{B}\ (\times10^{-3})$   &$\alpha$ \\
\midrule[0.5pt]
$\Xi_b^{0,-}\to\Xi_c^{+,0}\pi^-$         &4.04 (4.29)                      &-1.000
&$\Xi_b^{0,-}\to\Xi_c^{+,0}\rho^-$        &13.3 (14.1)                      &-0.792   \\
$\Xi_b^{0,-}\to\Xi_c^{+,0}K^-$           &0.31 (0.33)                      &-1.000
&$\Xi_b^{0,-}\to\Xi_c^{+,0}K^{*-}$        &0.71 (0.76)                      &-0.737   \\
$\Xi_b^{0,-}\to\Xi_c^{+,0}D^-$           &0.58 (0.62)                      &-0.983
&$\Xi_b^{0,-}\to\Xi_c^{+,0}D^{*-}$        &1.51 (1.60)                      &-0.239    \\
$\Xi_b^{0,-}\to\Xi_c^{+,0}D_s^-$         &14.8 (15.7)                      &-0.978
&$\Xi_b^{0,-}\to\Xi_c^{+,0}D_s^{*-}$      &32.4 (34.4)                      &-0.206    \\
$\Xi_b^{0,-}\to\Xi_c^{+,0}(2970)\pi^-$   &1.78 (1.89)                      &-0.999
&$\Xi_b^{0,-}\to\Xi_c^{+,0}(2970)\rho^-$  &2.78 (2.95)                      &-0.763   \\
$\Xi_b^{0,-}\to\Xi_c^{+,0}(2970)K^-$     &0.04 (0.05)                      &-0.998
&$\Xi_b^{0,-}\to\Xi_c^{+,0}(2970)K^{*-}$  &0.09 (0.10)                      &-0.702   \\
$\Xi_b^{0,-}\to\Xi_c^{+,0}(2970)D^-$     &0.04 (0.05)                      &-0.952
&$\Xi_b^{0,-}\to\Xi_c^{+,0}(2970)D^{*-}$  &0.12 (0.12)                      &-0.181    \\
$\Xi_b^{0,-}\to\Xi_c^{+,0}(2970)D_s^-$   &1.05 (1.12)                      &-0.940
&$\Xi_b^{0,-}\to\Xi_c^{+,0}(2970)D_s^{*-}$&2.30 (2.45)                      &-0.148    \\
\bottomrule[0.5pt]
\bottomrule[0.6pt]
\end{tabular*}
\end{table*}

\begin{table*}[htbp]\centering
\caption{The branching rates and asymmetry parameters of $\Omega_b\to\Omega_c^{(*)}M$ transitions with $M$ denoting a pseudoscalar or vector pmeson.}
\label{tab:nonleptonOmega}
\renewcommand\arraystretch{1.2}
\begin{tabular*}{172mm}{l@{\extracolsep{\fill}}ccl@{\extracolsep{\fill}}cc}
\toprule[0.5pt]
\toprule[0.5pt]
Mode                                       &$\mathcal{B}\ (\times10^{-3})$     &$\alpha$
&Mode                                      &$\mathcal{B}\ (\times10^{-3})$    &$\alpha$\\
\midrule[0.5pt]
$\Omega_b^{-}\to\Omega_c^{0}\pi^-$         &2.82                               &0.59
&$\Omega_b^{-}\to\Omega_c^{0}\rho^-$        &7.92                       &0.61    \\
$\Omega_b^{-}\to\Omega_c^{0}K^-$           &0.22                               &0.58
&$\Omega_b^{-}\to\Omega_c^{0}K^{*-}$        &0.41                       &0.62  \\
$\Omega_b^{-}\to\Omega_c^{0}D^-$           &0.52                               &0.49
&$\Omega_b^{-}\to\Omega_c^{0}D^{*-}$        &0.48                       &0.69   \\
$\Omega_b^{-}\to\Omega_c^{0}D_s^-$         &13.5                               &0.47
&$\Omega_b^{-}\to\Omega_c^{0}D_s^{*-}$      &9.73                       &0.70   \\
$\Omega_b^{-}\to\Omega_c^{0}(2S)\pi^-$   &0.30                               &0.58
&$\Omega_b^{-}\to\Omega_c^{0}(2S)\rho^-$       &0.70                  &0.60    \\
$\Omega_b^{-}\to\Omega_c^{0}(2S)K^-$     &0.02                               &0.57
&$\Omega_b^{-}\to\Omega_c^{0}(2S)K^{*-}$       &0.03                  &0.60   \\
$\Omega_b^{-}\to\Omega_c^{0}(2S)D^-$     &0.03                               &0.45
&$\Omega_b^{-}\to\Omega_c^{0}(2S)D^{*-}$       &0.02                  &0.65  \\
$\Omega_b^{-}\to\Omega_c^{0}(2S)D_s^-$   &0.62                               &0.43
&$\Omega_b^{-}\to\Omega_c^{0}(2S)D_s^{*-}$     &0.36                  &0.65   \\
\bottomrule[0.5pt]
\bottomrule[0.6pt]
\end{tabular*}
\end{table*}

\begin{table*}[htbp]\centering
\caption{Comparison of theoretical predictions for $\mathcal{B}(\Xi_b^{0,-}\to\Xi_c^{+,0}M^-)$ and $\mathcal{B}(\Omega_b^-\to\Omega_c^0M^-)$. Here, all values should be multiplied by a factor of $10^{-3}$.}
\label{tab:Comparison}
\renewcommand\arraystretch{1.2}
\begin{tabular*}{172mm}{l@{\extracolsep{\fill}}cccccc}
\toprule[0.5pt]
\toprule[0.5pt]
                                    &This work     &Cheng \cite{Cheng:1996cs}    &Ivanov {\it et al.} \cite{Ivanov:1997hi,Ivanov:1997ra}    &Zhao \cite{Zhao:2018zcb}  &Gutsche {\it et al.} \cite{Gutsche:2018utw}  &Chua \cite{Chua:2019yqh}  \\
\midrule[0.5pt]
$\Xi_b^{0,-}\to\Xi_c^{+,0}\pi^-$    &4.03\ (4.29) &4.9\ (5.2)  &7.08\ (10.13)   &8.37\ (8.93)   &$-$   &$3.66^{+2.29}_{-1.59}$\ ($3.88^{+2.43}_{-1.69}$)     \\
$\Xi_b^{0,-}\to\Xi_c^{+,0}\rho^-$   &13.3\ (14.1) &$-$         &$-$             &24.0\ (25.6)   &$-$   &$10.88^{+6.83}_{-4.74}$\ ($11.56^{+7.25}_{-5.04}$)   \\
$\Xi_b^{0,-}\to\Xi_c^{+,0}K^-$      &0.31\ (0.33) &$-$         &$-$             &0.667\ (0.711) &$-$   &$0.28^{+0.17}_{-0.12}$\ ($0.29^{+0.18}_{-0.13}$)     \\
$\Xi_b^{0,-}\to\Xi_c^{+,0}K^{*-}$   &0.71\ (0.76) &$-$         &$-$             &1.23\ (1.31)   &$-$   &$0.56^{+0.35}_{-0.24}$\ ($0.60^{+0.37}_{-0.26}$)     \\
$\Xi_b^{0,-}\to\Xi_c^{+,0}D^-$      &0.58\ (0.62) &$-$         &$-$             &0.949\ (1.03)  &0.45  &$0.43^{+0.29}_{-0.20}$\ ($0.45^{+0.31}_{-0.21}$)     \\
$\Xi_b^{0,-}\to\Xi_c^{+,0}D^{*-}$   &1.51\ (1.60) &$-$         &$-$             &1.54\ (1.64)   &0.95  &$0.77^{+0.50}_{-0.35}$\ ($0.82^{+0.53}_{-0.37}$)     \\
$\Xi_b^{0,-}\to\Xi_c^{+,0}D_s^-$    &14.8\ (15.7) &14.6        &$-$             &24.6\ (26.2)   &$-$   &$10.87^{+7.51}_{-5.03}$\ ($11.54^{+7.98}_{-5.34}$)   \\
$\Xi_b^{0,-}\to\Xi_c^{+,0}D_s^{*-}$ &32.4\ (34.4) &23.1        &$-$             &36.5\ (39.0)   &$-$   &$16.24^{+10.54}_{-7.25}$\ ($17.26^{+11.2}_{-7.70}$)  \\
$\Omega_b^-\to\Omega_c^0\pi^-$      &2.82         &4.92        &5.81            &4.00           &1.88  &$1.10^{+0.85}_{-0.55}$                               \\
$\Omega_b^-\to\Omega_c^0\rho^-$     &7.92         &12.8        &$-$             &10.8           &5.43  &$3.07^{+2.41}_{-1.53}$                               \\
$\Omega_b^-\to\Omega_c^0K^-$        &0.22         &$-$         &$-$             &0.326          &$-$   &$0.08^{+0.07}_{-0.04}$                               \\
$\Omega_b^-\to\Omega_c^0K^{*-}$     &0.41         &$-$         &$-$             &0.544          &$-$   &$0.16^{+0.12}_{-0.08}$                               \\
$\Omega_b^-\to\Omega_c^0D^-$        &0.52         &$-$         &$-$             &0.636          &$-$   &$0.15^{+0.14}_{-0.08}$                               \\
$\Omega_b^-\to\Omega_c^0D^{*-}$     &0.48         &$-$         &$-$             &0.511          &$-$   &$0.16^{+0.13}_{-0.08}$                               \\
$\Omega_b^-\to\Omega_c^0D_s^-$      &13.5         &17.9        &$-$             &17.1           &$-$   &$4.03^{+3.72}_{-2.21}$                               \\
$\Omega_b^-\to\Omega_c^0D_s^{*-}$   &9.73         &11.5        &$-$             &11.7           &$-$   &$3.18^{+2.69}_{-1.61}$                               \\
\bottomrule[0.5pt]
\bottomrule[0.5pt]
\end{tabular*}
\end{table*}

\section{Summary}
\label{sec5}

With the accumulation of experimental data from LHCb and Belle II \cite{Belle-II:2018jsg},
experimental exploration of weak decay of the bottom baryons $\Xi_b$ and $\Omega_b$ is becoming possible. Facing this opportunity, in this work we study the color-allowed two-body nonleptonic decay of the bottom baryons $\Xi_b$ and $\Omega_b$, i.e., the $\Xi_b\to \Xi_c^{(*)}M$ and $\Omega_b\to \Omega_c^{(*)}M$ decay with emitting a pseudoscalar meson ($\pi^{-}$, $K^{-}$, $D^{-}$, and $D_s^{-}$) or a vector meson ($\rho^{-}$, $K^{*-}$, $D^{*-}$, and $D_s^{*-}$).

We adopt the three-body light-front quark model to calculate these form factors depicting these discussed bottom baryon to the charmed baryon transitions under the na\"ive factorization framework. We also improve the treatment of the spatial wave function of these involved heavy baryons in these decays, where the semirelativistic three-body potential model \cite{Capstick:1985xss,Li:2021qod} is applied to calculate the numerical spatial wave function of these heavy baryons with the help of the GEM \cite{Hiyama:2003cu,Yoshida:2015tia,Hiyama:2018ivm,Yang:2019lsg}. We call that the study of color-allowed two-body nonleptonic decay of bottom baryons $\Xi_b$ and $\Omega_b$ is supported by hadron spectroscopy.
Our result shows that these color-allowed two-body nonleptonic decays $\Xi_b^{0,-}\to\Xi_c^{(*)+,0}$ and $\Omega_b^{-}\to\Omega_c^{(*)0}$ with the $\pi^-$, $\rho^-$, and $D_s^{(*)-}$-emitted modes have considerable widths.

We suggest to measure these discussed color-allowed two-body nonleptonic decay of the bottom baryons $\Xi_b$ and $\Omega_b$, which will be good chance for the ongoing LHCb and Belle II experiments.

\section*{ACKNOWLEDGMENTS}

This work is supported by the China National Funds for Distinguished Young Scientists under Grant No. 11825503, National Key Research and Development Program of China under Contract No. 2020YFA0406400, the 111 Project under Grant No. B20063, the National Natural Science Foundation of China under Grant No. 12047501, and by the Fundamental Research Funds for the Central Universities.


\begin{thebibliography}{99}

\bibitem{BaBar:2012obs}
J.~P.~Lees \textit{et al.} [BaBar],
Evidence for an excess of $\bar{B} \to D^{(*)} \tau^-\bar{\nu}_\tau$ decays,
Phys. Rev. Lett. \textbf{109} (2012), 101802.

\bibitem{BaBar:2013mob}
J.~P.~Lees \textit{et al.} [BaBar],
Measurement of an Excess of $\bar{B} \to D^{(*)}\tau^- \bar{\nu}_\tau$ Decays and Implications for Charged Higgs Bosons,
Phys. Rev. D \textbf{88} (2013) no.7, 072012.

\bibitem{Belle:2015qfa}
M.~Huschle \textit{et al.} [Belle],
Measurement of the branching ratio of $\bar{B} \to D^{(\ast)} \tau^- \bar{\nu}_\tau$ relative to $\bar{B} \to D^{(\ast)} \ell^- \bar{\nu}_\ell$ decays with hadronic tagging at Belle,
Phys. Rev. D \textbf{92} (2015) no.7, 072014.

\bibitem{LHCb:2015gmp}
R.~Aaij \textit{et al.} [LHCb],
Measurement of the ratio of branching fractions $\mathcal{B}(\bar{B}^0 \to D^{*+}\tau^{-}\bar{\nu}_{\tau})/\mathcal{B}(\bar{B}^0 \to D^{*+}\mu^{-}\bar{\nu}_{\mu})$,
Phys. Rev. Lett. \textbf{115} (2015) no.11, 111803
[erratum: Phys. Rev. Lett. \textbf{115} (2015) no.15, 159901].

\bibitem{Belle:2016dyj}
S.~Hirose \textit{et al.} [Belle],
Measurement of the $\tau$ lepton polarization and $R(D^*)$ in the decay $\bar{B} \to D^* \tau^- \bar{\nu}_\tau$,
Phys. Rev. Lett. \textbf{118} (2017) no.21, 211801.

\bibitem{Belle:2019rba}
G.~Caria \textit{et al.} [Belle],
Measurement of $\mathcal{R}(D)$ and $\mathcal{R}(D^*)$ with a semileptonic tagging method,
Phys. Rev. Lett. \textbf{124} (2020) no.16, 161803.

\bibitem{FermilabLattice:2021cdg}
A.~Bazavov \textit{et al.} [Fermilab Lattice and MILC],
Semileptonic form factors for $B \to D^\ast\ell\nu$ at nonzero recoil from 2 + 1-flavor lattice QCD,
[arXiv:2105.14019 [hep-lat]].

\bibitem{HFLAV:2019otj}
Y.~S.~Amhis \textit{et al.} [HFLAV],
Averages of b-hadron, c-hadron, and $\tau $-lepton properties as of 2018,
Eur. Phys. J. C \textbf{81} (2021) no.3, 226.

\bibitem{CDF:2008llm}
T.~Aaltonen \textit{et al.} [CDF],
Observation of New Charmless Decays of Bottom Hadrons,
Phys. Rev. Lett. \textbf{103} (2009), 031801.

\bibitem{LHCb:2014yin}
R.~Aaij \textit{et al.} [LHCb],
Searches for $\Lambda^0_{b}$ and $\Xi^{0}_{b}$ decays to $K^0_{\rm S} p \pi^{-}$ and $K^0_{\rm S}p K^{-}$ final states with first observation of the $\Lambda^0_{b} \rightarrow K^0_{\rm S}p \pi^{-}$ decay,
JHEP \textbf{04} (2014), 087.

\bibitem{LHCb:2016rja}
R.~Aaij \textit{et al.} [LHCb],
Observations of $\Lambda_b^0 \to \Lambda K^+\pi^-$ and $\Lambda_b^0 \to \Lambda K^+K^-$ decays and searches for other $\Lambda_b^0$ and $\Xi_b^0$ decays to $\Lambda h^+h^{\prime -}$ final states,
JHEP \textbf{05} (2016), 081.

\bibitem{ParticleDataGroup:2020ssz}
P.~A.~Zyla \textit{et al.} [Particle Data Group],
Review of Particle Physics,
PTEP \textbf{2020} (2020) no.8, 083C01.

\bibitem{LHCb:2015yax}
R.~Aaij \textit{et al.} [LHCb],
Observation of $J/\psi p$ Resonances Consistent with Pentaquark States in $\Lambda_b^0 \to J/\psi K^- p$ Decays,
Phys. Rev. Lett. \textbf{115} (2015), 072001.

\bibitem{LHCb:2019kea}
R.~Aaij \textit{et al.} [LHCb],
Observation of a narrow pentaquark state, $P_c(4312)^+$, and of two-peak structure of the $P_c(4450)^+$,
Phys. Rev. Lett. \textbf{122} (2019) no.22, 222001.

\bibitem{LHCb:2020jpq}
R.~Aaij \textit{et al.} [LHCb],
Evidence of a $J/\psi\Lambda$ structure and observation of excited $\Xi^-$ states in the $\Xi^-_b \to J/\psi\Lambda K^-$ decay,
Sci. Bull. \textbf{66} (2021), 1278-1287.

\bibitem{Belle-II:2018jsg}
E.~Kou \textit{et al.} [Belle-II],
The Belle II Physics Book,
PTEP \textbf{2019} (2019) no.12, 123C01
[erratum: PTEP \textbf{2020} (2020) no.2, 029201].

\bibitem{Faustov:2018ahb}
R.~N.~Faustov and V.~O.~Galkin,
Relativistic description of the $\Xi_b$ baryon semileptonic decays,
Phys. Rev. D \textbf{98} (2018) no.9, 093006.

\bibitem{Geng:2020ofy}
C.~Q.~Geng, C.~W.~Liu and T.~H.~Tsai,
Nonleptonic two-body weak decays of $\Lambda_b$ in modified MIT bag model,
Phys. Rev. D \textbf{102} (2020) no.3, 034033.

\bibitem{Albertus:2004wj}
C.~Albertus, E.~Hernandez and J.~Nieves,
Nonrelativistic constituent quark model and HQET combined study of semileptonic decays of $\Lambda_b$ and $\Xi_b$ baryons,
Phys. Rev. D \textbf{71} (2005), 014012.

\bibitem{Ebert:2006rp}
D.~Ebert, R.~N.~Faustov and V.~O.~Galkin,
Semileptonic decays of heavy baryons in the relativistic quark model,
Phys. Rev. D \textbf{73} (2006), 094002.

\bibitem{Cheng:1996cs}
H.~Y.~Cheng,
Nonleptonic weak decays of bottom baryons,
Phys. Rev. D \textbf{56} (1997), 2799-2811
[erratum: Phys. Rev. D \textbf{99} (2019) no.7, 079901].

\bibitem{Ivanov:1997hi}
M.~A.~Ivanov, J.~G.~Korner, V.~E.~Lyubovitskij and A.~G.~Rusetsky,
Exclusive nonleptonic bottom to charm baryon decays including nonfactorizable contributions,
Mod. Phys. Lett. A \textbf{13} (1998), 181-192.

\bibitem{Ivanov:1997ra}
M.~A.~Ivanov, J.~G.~Korner, V.~E.~Lyubovitskij and A.~G.~Rusetsky,
Exclusive nonleptonic decays of bottom and charm baryons in a relativistic three quark model: Evaluation of nonfactorizing diagrams,
Phys. Rev. D \textbf{57} (1998), 5632-5652.

\bibitem{Gutsche:2018utw}
T.~Gutsche, M.~A.~Ivanov, J.~G.~K\"orner and V.~E.~Lyubovitskij,
Nonleptonic two-body decays of single heavy baryons  $\Lambda_Q$, $\Xi_Q$, and $\Omega_Q$ $(Q=b,c)$ induced by $W$ emission in the covariant confined quark model,
Phys. Rev. D \textbf{98} (2018) no.7, 074011.

\bibitem{Zhao:2018zcb}
Z.~X.~Zhao,
Weak decays of heavy baryons in the light-front approach,
Chin. Phys. C \textbf{42} (2018) no.9, 093101.

\bibitem{Chua:2018lfa}
C.~K.~Chua,
Color-allowed bottom baryon to charmed baryon nonleptonic decays,
Phys. Rev. D \textbf{99} (2019) no.1, 014023.

\bibitem{Chua:2019yqh}
C.~K.~Chua,
Color-allowed bottom baryon to $s$-wave and $p$-wave charmed baryon nonleptonic decays,
Phys. Rev. D \textbf{100} (2019) no.3, 034025.

\bibitem{Ke:2019smy}
H.~W.~Ke, N.~Hao and X.~Q.~Li,
Revisiting $\Lambda _{b}\rightarrow \Lambda _{c}$ and $\Sigma _{b}\rightarrow \Sigma _{c}$ weak decays in the light-front quark model,
Eur. Phys. J. C \textbf{79} (2019) no.6, 540.

\bibitem{Zhu:2018jet}
J.~Zhu, Z.~T.~Wei and H.~W.~Ke,
Semileptonic and nonleptonic weak decays of $\Lambda_b^0$,
Phys. Rev. D \textbf{99} (2019) no.5, 054020.

\bibitem{Li:2021qod}
Y.~S.~Li, X.~Liu and F.~S.~Yu,
Revisiting semileptonic decays of \ensuremath{\Lambda}b(c) supported by baryon spectroscopy,
Phys. Rev. D \textbf{104} (2021) no.1, 013005.

\bibitem{Wang:2008sm}
Y.~m.~Wang, Y.~Li and C.~D.~Lu,
Rare Decays of $\Lambda_b\to\Lambda+\gamma$ and $\Lambda_b\to\Lambda\ell^+\ell^-$ in the Light-cone Sum Rules,
Eur. Phys. J. C \textbf{59} (2009), 861-882.

\bibitem{Khodjamirian:2011jp}
A.~Khodjamirian, C.~Klein, T.~Mannel and Y.~M.~Wang,
Form Factors and Strong Couplings of Heavy Baryons from QCD Light-Cone Sum Rules,
JHEP \textbf{09} (2011), 106.

\bibitem{Wang:2015ndk}
Y.~M.~Wang and Y.~L.~Shen,
Perturbative Corrections to $\Lambda_b \to \Lambda$ Form Factors from QCD Light-Cone Sum Rules,
JHEP \textbf{02} (2016), 179.

\bibitem{Zhao:2020mod}
Z.~X.~Zhao, R.~H.~Li, Y.~L.~Shen, Y.~J.~Shi and Y.~S.~Yang,
The semi-leptonic form factors of $\Lambda_{b}\to\Lambda_{c}$ and $\Xi_{b}\to\Xi_{c}$ in QCD sum rules,
Eur. Phys. J. C \textbf{80} (2020) no.12, 1181.

\bibitem{Guo:2005qa}
P.~Guo, H.~W.~Ke, X.~Q.~Li, C.~D.~Lu and Y.~M.~Wang,
Diquarks and the semi-leptonic decay of $\Lambda_b$ in the hyrid scheme,
Phys. Rev. D \textbf{75} (2007), 054017.

\bibitem{Capstick:1985xss}
S.~Capstick and N.~Isgur,
Baryons in a Relativized Quark Model with Chromodynamics,
AIP Conf. Proc. \textbf{132} (1985), 267-271.

\bibitem{Hiyama:2018ivm}
E.~Hiyama and M.~Kamimura,
Study of various few-body systems using Gaussian expansion method (GEM),
Front. Phys. (Beijing) \textbf{13} (2018) no.6, 132106.

\bibitem{Hiyama:2003cu}
E.~Hiyama, Y.~Kino and M.~Kamimura,
Gaussian expansion method for few-body systems,
Prog. Part. Nucl. Phys. \textbf{51} (2003), 223-307.

\bibitem{Yoshida:2015tia}
T.~Yoshida, E.~Hiyama, A.~Hosaka, M.~Oka and K.~Sadato,
Spectrum of heavy baryons in the quark model,
Phys. Rev. D \textbf{92} (2015) no.11, 114029.

\bibitem{Yang:2019lsg}
G.~Yang, J.~Ping, P.~G.~Ortega and J.~Segovia,
Triply heavy baryons in the constituent quark model,
Chin. Phys. C \textbf{44} (2020) no.2, 023102.

\bibitem{Tawfiq:1998nk}
S.~Tawfiq, P.~J.~O'Donnell and J.~G.~Korner,
Charmed baryon strong coupling constants in a light front quark model,
Phys. Rev. D \textbf{58} (1998), 054010.

\bibitem{Godfrey:1985xj}
S.~Godfrey and N.~Isgur,
Mesons in a Relativized Quark Model with Chromodynamics,
Phys. Rev. D \textbf{32} (1985), 189-231.

\bibitem{Song:2015nia}
Q.~T.~Song, D.~Y.~Chen, X.~Liu and T.~Matsuki,
Charmed-strange mesons revisited: mass spectra and strong decays,
Phys. Rev. D \textbf{91} (2015), 054031.

\bibitem{Pang:2017dlw}
C.~Q.~Pang, J.~Z.~Wang, X.~Liu and T.~Matsuki,
A systematic study of mass spectra and strong decay of strange mesons,
Eur. Phys. J. C \textbf{77} (2017) no.12, 861.

\bibitem{Wang:2018rjg}
J.~Z.~Wang, Z.~F.~Sun, X.~Liu and T.~Matsuki,
Higher bottomonium zoo,
Eur. Phys. J. C \textbf{78} (2018) no.11, 915.

\bibitem{Wang:2019mhs}
J.~Z.~Wang, D.~Y.~Chen, X.~Liu and T.~Matsuki,
Constructing $J/\psi$ family with updated data of charmoniumlike $Y$ states,
Phys. Rev. D \textbf{99} (2019) no.11, 114003.

\bibitem{Duan:2021alw}
M.~X.~Duan and X.~Liu,
Where are 3P and higher P-wave states in the charmonium family?,
Phys. Rev. D \textbf{104} (2021) no.7, 074010.

\bibitem{Chen:2007xf}
C.~Chen, X.~L.~Chen, X.~Liu, W.~Z.~Deng and S.~L.~Zhu,
Strong decays of charmed baryons,
Phys. Rev. D \textbf{75} (2007), 094017.

\bibitem{Chen:2021eyk}
B.~Chen, S.~Q.~Luo and X.~Liu,
Universal behavior of mass gaps existing in the single heavy baryon family,
Eur. Phys. J. C \textbf{81} (2021) no.5, 474.

\bibitem{Biagi:1983en}
S.~F.~Biagi, M.~Bourquin, A.~J.~Britten, R.~M.~Brown, H.~J.~Burckhart, A.~A.~Carter, C.~Dore, P.~Extermann, M.~Gailloud and C.~N.~P.~Gee, \textit{et al.}
Observation of a Narrow State at $2.46\ \text{GeV}/c^2$: A Candidate for the Charmed Strange Baryon $A^+$,
Phys. Lett. B \textbf{122} (1983), 455.

\bibitem{FermilabE687:1992wmm}
P.~L.~Frabetti \textit{et al.} [Fermilab E687],
Measurement of the mass and lifetime of the $\Xi_c^+$,
Phys. Rev. Lett. \textbf{70} (1993), 1381-1384.

\bibitem{CLEO:1988yda}
P.~Avery \textit{et al.} [CLEO],
Observation of the Charmed Strange Baryon $\Xi_c^0$,
Phys. Rev. Lett. \textbf{62} (1989), 863.

\bibitem{Belle:2006edu}
R.~Chistov \textit{et al.} [Belle],
Observation of new states decaying into $\Lambda_c^+K^-\pi^+$ and $\Lambda_c^+\to K_S^0\pi^-$,
Phys. Rev. Lett. \textbf{97} (2006), 162001.

\bibitem{Biagi:1984mu}
S.~F.~Biagi, M.~Bourquin, A.~J.~Britten, R.~M.~Brown, H.~J.~Burckhart, A.~A.~Carter, C.~Dor\'e, P.~Extermann, M.~Gailloud and C.~N.~P.~Gee, \textit{et al.}
Properties of the Charmed Strange Baryon A+ and Evidence for the Charmed Doubly Strange Baryon $T^0$ at $2.74\ \text{GeV}/c^2$,
Z. Phys. C \textbf{28} (1985), 175.

\bibitem{ARGUS:1992mwl}
H.~Albrecht \textit{et al.} [ARGUS],
Evidence for the production of the charmed, doubly strange baryon $\Omega_c$ in $e^+e^-$ annihilation,
Phys. Lett. B \textbf{288} (1992), 367-372.

\bibitem{LHCb:2017uwr}
R.~Aaij \textit{et al.} [LHCb],
Observation of five new narrow $\Omega_c^0$ states decaying to $\Xi_c^+ K^-$,
Phys. Rev. Lett. \textbf{118} (2017) no.18, 182001.

\bibitem{Belle:2017ext}
J.~Yelton \textit{et al.} [Belle],
Observation of Excited $\Omega_c$ Charmed Baryons in $e^+e^-$ Collisions,
Phys. Rev. D \textbf{97} (2018) no.5, 051102.

\bibitem{Chen:2017gnu}
B.~Chen and X.~Liu,
New $\Omega_c^0$ baryons discovered by LHCb as the members of $1P$ and $2S$ states,
Phys. Rev. D \textbf{96} (2017) no.9, 094015.

\bibitem{Cheng:2017ove}
H.~Y.~Cheng and C.~W.~Chiang,
Quantum numbers of $\Omega_c$ states and other charmed baryons,
Phys. Rev. D \textbf{95} (2017) no.9, 094018.

\bibitem{Chen:2017sci}
H.~X.~Chen, Q.~Mao, W.~Chen, A.~Hosaka, X.~Liu and S.~L.~Zhu,
Decay properties of $P$-wave charmed baryons from light-cone QCD sum rules,
Phys. Rev. D \textbf{95} (2017) no.9, 094008.

\bibitem{Agaev:2017jyt}
S.~S.~Agaev, K.~Azizi and H.~Sundu,
On the nature of the newly discovered $\Omega$ states,
EPL \textbf{118} (2017) no.6, 61001.

\bibitem{Wang:2017hej}
K.~L.~Wang, L.~Y.~Xiao, X.~H.~Zhong and Q.~Zhao,
Understanding the newly observed $\Omega_c$ states through their decays,
Phys. Rev. D \textbf{95} (2017) no.11, 116010.

\bibitem{Debastiani:2018adr}
V.~R.~Debastiani, J.~M.~Dias, W.~H.~Liang and E.~Oset,
$\Omega_b^- \to (\Xi_c^+ \, K^-) \, \pi^-$ and the $\Omega_c$ states,
Phys. Rev. D \textbf{98} (2018) no.9, 094022.

\bibitem{Georgi:1990ei}
H.~Georgi, B.~Grinstein and M.~B.~Wise,
$\Lambda_b$ semileptonic decay form-factors for $m_c$ does not equal infinity,
Phys. Lett. B \textbf{252} (1990), 456-460.

\bibitem{Bowler:1997ej}
K.~C.~Bowler \textit{et al.} [UKQCD],
First lattice study of semileptonic decays of $\Lambda_b$ and $\Xi_b$ baryons,
Phys. Rev. D \textbf{57} (1998), 6948-6974.

\bibitem{Lu:2009cm}
C.~D.~Lu, Y.~M.~Wang, H.~Zou, A.~Ali and G.~Kramer,
Anatomy of the pQCD Approach to the Baryonic Decays $\Lambda_b\to p\pi, pK$,
Phys. Rev. D \textbf{80} (2009), 034011.

\bibitem{LHCb:2014chk}
R.~Aaij \textit{et al.} [LHCb],
Precision measurement of the mass and lifetime of the $\Xi_b^0$ baryon,
Phys. Rev. Lett. \textbf{113} (2014), 032001.

\bibitem{LHCb:2014wqn}
R.~Aaij \textit{et al.} [LHCb],
Measurement of the $\Xi_b^-$ and $\Omega_b^-$ baryon lifetimes,
Phys. Lett. B \textbf{736} (2014), 154-162.

\bibitem{LHCb:2016coe}
R.~Aaij \textit{et al.} [LHCb],
Measurement of the mass and lifetime of the $\Omega_b^-$ baryon,
Phys. Rev. D \textbf{93} (2016) no.9, 092007.

\bibitem{CDF:2014mon}
T.~A.~Aaltonen \textit{et al.} [CDF],
Mass and lifetime measurements of bottom and charm baryons in $p\bar p$ collisions at $\sqrt{s}= 1.96$ TeV,
Phys. Rev. D \textbf{89} (2014) no.7, 072014.

\bibitem{Cheng:2003sm}
H.~Y.~Cheng, C.~K.~Chua and C.~W.~Hwang,
Covariant light front approach for $S$ wave and $P$ wave mesons: Its application to decay constants and form-factors,
Phys. Rev. D \textbf{69} (2004), 074025.

\end{thebibliography}
\end{document}